\documentclass[12pt]{article}
\usepackage[
top = 2.5cm, 
bottom = 2.5cm, 
left = 2.5cm, 
right = 2.5cm]{geometry}

\usepackage{amssymb,amsfonts,amsmath,amssymb,amscd,amstext,color}
\usepackage{jheppub}
\usepackage{amsthm}
\usepackage{slashed}
\usepackage{color}
\usepackage{graphicx}
\usepackage{hyperref}

\numberwithin{equation}{section}

\numberwithin{problem}{section}

\def\be{\begin{eqnarray}}
\def\ee{\end{eqnarray}}

\def\parscalar{\hat \pi}
\def\perpscalar{\hat \tau}
\def\parscalarnohat{\pi}
\def\perpscalarnohat{\tau}

\title{
Conformal Surface Defects in Maxwell Theory are Trivial
}
\author{Christopher P. Herzog and Abhay Shrestha}
\affiliation{Mathematics Department, King's College London, \\
The Strand, London,  WC2R 2LS, UK }
\abstract{
We consider a free Maxwell field in four dimensions in the presence of a codimension two defect.
Reflection positive, codimension two defects which preserve conformal symmetry in this context are very limited. 
We show only generalized free fields can appear in the defect operator product expansion of the bulk Maxwell field;
in particular correlation functions of these defect operators can be evaluated via Wick's Theorem.
}

\begin{document}
\maketitle

\section{Introduction}

We are interested in finding computable models of boundary and defect conformal field theories.  
These are $d$-dimensional quantum field theories where a $p$-dimensional defect reduces the 
(Euclidean) conformal symmetry group from
$SO(d+1,1)$ to $SO(p+1,1) \times SO(d-p)$.  These types of field theories find applications in a variety of contexts, from 
critical systems in condensed matter, to entanglement phenomena in many-body physics, to D-branes in string theory. 
One of us
\cite{Herzog:2017xha} proposed that theories with interactions confined to the boundary but that were free in the bulk
might be a nontrivial yet tractable set to investigate.  Since that proposal, there has been an enormous amount of work
 that has greatly illuminated the situation, both by finding new results
 and also by uncovering relations to earlier research, in some cases decades
 earlier.

One of the most interesting free-in-the-bulk theories proposed in \cite{Herzog:2017xha}, sometimes called reduced quantum electrodynamics (reduced QED),
 was a free Maxwell field in four dimensions
interacting with charged matter confined to a three dimensional surface. 
 In this case, a combination of gauge invariance and bulk locality
sets the beta function for the electromagnetic coupling to zero, as was first noticed perturbatively
\cite{Teber:2012de,Kotikov:2013eha} and then later argued to hold at all loops \cite{Herzog:2017xha, Dudal:2018pta}.
The theory is interesting in part because of a rich history that predates the publication of  \cite{Herzog:2017xha}.
A careful treatment can be found already in \cite{Gorbar:2001qt} in the context of dynamical symmetry breaking and spontaneous
mass generation for fermions. 
The theory also has a close relation to the large $N$ limit of three dimensional QED \cite{Kotikov:2016yrn}.  
Probably the most interesting fact about this mixed dimensional QED theory
 is that, in an idealized sense, it is the infrared renormalization group fixed point of graphene
\cite{Kotikov:2013eha}.  See \cite{Olivares:2021svj} for a recent in depth discussion of this relation, and a more careful treatment of the history.

At this point, reduced QED is the poster child for computable models of boundary and defect CFT that are free in the bulk.
$SL(2, {\mathbb Z})$ duality allows one access to strong coupling \cite{Hsiao:2017lch,Hsiao:2018fsc,DiPietro:2019hqe}.  
There are supersymmetric versions, dubbed supergraphene \cite{Herzog:2018lqz,James:2021ggq,KumarGupta:2019nay,Gupta:2020eev}.  
Indeed, the ${\mathcal N}=2$ supersymmetric version allows for a localization approach where one can compute
transport quantities exactly at any value of the coupling \cite{KumarGupta:2019nay,Gupta:2020eev}.

Theories with free scalars or fermions in the bulk have thus far presented additional difficulties.  The main issue is that the boundary couplings
tend to run, and the fixed points are not under fine control.  The earliest example we found was from 35 years ago \cite{DiehlEisenrieglerArticle},
where a free scalar field in three dimensions interacts via a classically marginal $\phi^4$ coupling on a two dimensional boundary.
More recent work utilizing $\epsilon$ and large $N$ expansions to study boundary interactions
includes \cite{Prochazka:2020vog,Giombi:2019enr,Giombi:2020rmc,DiPietro:2020fya,Giombi:2021cnr,Herzog:2020lel}.

We have focused in this discussion thus far on the codimension one or boundary case.  It is natural to wonder what happens then in higher codimension.
An important recent result \cite{Lauria:2020emq} shows that scalar field theories that are free in the bulk must be ``trivial'' in codimension higher than one. 
``Trivial'' here means that all of the defect operators that have a nonzero three point function with the bulk scalar must satisfy a
``double twist'' condition on their spectrum.  Moreover, correlation functions involving fields that appear in the defect operator product
expansion (OPE) of the scalar can be computed from their two-point functions via Wick's Theorem.  ``Double twist'' means that if two of the operators in the three-point function are $\hat {\mathcal O}_1$ and $\hat {\mathcal O}_2$, and at least one of $\hat {\mathcal O}_1$ or $\hat {\mathcal O}_2$ appears in the defect OPE of the scalar, the third operator must take the schematic form
$\hat {\mathcal O}_1 \partial^n \hat {\mathcal O}_2$.  
The authors restricted their result to scalars for simplicity.  The steps of their argument, however, can be applied more generally.
In this paper, we extend their result to the case of a Maxwell field in four dimensions
interacting with a two dimensional defect.  (We limit ourselves to the codimension two case because of the proliferation of tensor structures that appear in codimension three.)

There have been investigations of a Maxwell field in higher codimension interacting with charged matter on the defect 
\cite{Gorbar:2001qt,Heydeman:2020ijz}.
Indeed, from their work, it is already suggestive what goes wrong.  In perturbation theory, it is convenient to work in a partially Fourier
transformed setting where we replace the coordinates along the defect with their corresponding momenta.  However, having done that, the photon propagator
along the defect will have a logarithmic dependence on the momentum, $\log (p)$, requiring the introduction of a scale, even at tree level, in the computation
of Feynman diagrams.  
The task of the work below is to replace this perturbative argument with something more rigorous based purely on symmetry and unitarity.

Another output of the work here is a technical advance --  formalism for dealing with mixed symmetry correlation functions in defect CFT.
In \cite{Herzog:2020bqw}, developing the boundary CFT technology of \cite{McAvity:1993ue,McAvity:1995zd}, we proposed a method
for writing down two-point functions in defect CFT involving operators transforming in arbitrary representations of the Lorentz group.  In this paper,
we necessarily need to push that formalism a little further, in order to be able to write down a bulk-defect-defect three-point function involving a 
Maxwell field $F_{\mu\nu}$.  
There is a competing product on the market available for defect CFT, based on the embedding space formalism, that was developed in a series of papers
\cite{Billo:2016cpy,Lauria:2017wav,Lauria:2018klo} and later further elaborated \cite{Guha:2018snh,Kobayashi:2018okw}.
While we are fans of embedding space for symmetric representations of the Lorentz group, we prefer the methods here
for antisymmetric tensors such as $F_{\mu\nu}$.

The outline of the rest of this work is as follows.  In section \ref{sec:FF},  we calculate the $\langle F_{\mu\nu}(x) F_{\lambda \rho}(x') \rangle$ two point function  and make some side remarks about the stress tensor for orbifold theories.  
In section \ref{sec:FO}, we compute bulk-defect two-point functions involving the Maxwell field $F_{\mu\nu}$.  
In section \ref{sec:conformalblock}, we reconstruct the defect OPE of $F_{\mu\nu}$ from the bulk-defect two-point functions, and then re-express 
$\langle F_{\mu\nu}(x) F_{\lambda \rho}(x') \rangle$ as a sum over conformal blocks.  At the end of these three exercises, we discover that only vectors
of conformal weight $\Delta = 1+|s|$ and a pair of complex scalars of weight $\Delta =2$
 contribute to the defect OPE of $F_{\mu\nu}$.  Here $s$ is the transverse spin of the defect operators.
Section \ref{sec:FWW} computes the bulk-defect-defect three-point function and from it extracts severe restrictions on the spectrum of defect operators.
Section \ref{sec:trivial} uses these restrictions on the spectrum to prove the ``triviality'' of the theory.
We conclude with a short discussion, and various appendices contain auxiliary results.

\section{Field Strength Two Point Function}
\label{sec:FF}

In this section, we constrain the form of the $\langle F_{\mu\nu}(x) F_{\lambda \rho}(x') \rangle$ two-point function
 for a Maxwell theory in four dimensions
in the presence of a codimension two defect.  
While not strictly needed for the central arguments of the work, 
$\langle F_{\mu\nu}(x) F_{\lambda \rho}(x') \rangle$ gives us an opportunity to set up needed notation
and is also an important correlation function in the theory.
We assume that the defect breaks the $SO(5,1)$ conformal symmetry
of the (Euclidean) Maxwell theory to $SO(3,1) \times SO(2)$.
In ref.\ \cite{Herzog:2020bqw}, we developed a formalism for constructing precisely this type of correlation function.
  The formalism begins with
the construction of cross ratios which are invariant under the residual $SO(3,1) \times SO(2)$ conformal group.

To fix notation, 
we insert the two Maxwell fields at points $x$ and $x' \in {\mathbb R}^4$.  The defect sits at $y = 0$ where
$y \in {\mathbb R}^2$ is a point in the space transverse to the defect while $\textbf{x} \in \mathbb{R}^2$ is a point on the defect. The two cross ratios which took pride of place in ref.\ \cite{Herzog:2020bqw} are
\begin{equation} \label{crossratio}
\begin{split}
\xi_1 =\frac{(x-x')^2}{4 |y| |y'|}, 
\qquad
\xi_2 = \frac{y \cdot y'}{|y| |y'|}.
\end{split}
\end{equation}

To solve the free field constraints on $\langle F_{\mu\nu}(x) F_{\lambda \rho}(x') \rangle$, it is convenient to
alter the formalism \cite{Herzog:2020bqw} in a way that will allow us at a later stage to perform a separation of
variables into a radial coordinate $r$ and angular coordinate $\theta$.  
The cross ratio $\xi_2 = \cos \theta$  depends only on the angle between the vectors $y$ and $y'$.  
$\xi_1$ on the other hand depends on both $\theta$ and the magnitudes of $y$ and $y'$.  By
replacing $\xi_1$ with the quantity \cite{Lauria:2017wav}
\be
\chi = 2 \xi_1 + \xi_2 =
\frac{({\bf x} - {\bf x}')^2 + y^2 + y'^2}{2 |y| |y'| } \ ,
\ee
we find a new cross ratio which depends only on the magnitudes of $y$ and $y'$.  

The radial coordinate $r$ is then defined as
\be
\chi \equiv \frac{1+r^2}{2r}\ .
\ee  
This coordinate has a simple geometric interpretation. Splitting $x = (\textbf{x},y)$,
we use the conformal transformations to fix $\textbf{x} = \textbf{x}' = 0$ and $y' = (1,0)$.  The remaining two-vector $y$ we can write in complex notation
as $r e^{i \theta}$.  Note that the map from $r$ to $\chi$ is not injective: for every $r$ in the range $(0,1)$, there is a second $r>1$ that maps
to the same value of $\chi$.

In ref.\ \cite{Herzog:2020bqw},  we introduced tensor building blocks from which to construct the correlation functions.
Among the vectors, we have
\begin{equation} \label{confvectors}
\begin{aligned}
\Xi^{(1)}_{\mu}(x,x') &= \frac{|y|}{\xi_1} \frac{\partial \xi_1}{\partial x^{\mu}} 
, \\
\Xi^{(2)}_{\mu}(x,x') &=\frac{|y|}{\xi_2} \frac{\partial \xi_2}{\partial x^{\mu}}
, 
\end{aligned}
\qquad
\begin{aligned}
\Xi'^{(1)}_{\mu}(x,x') &= \frac{|y'|}{\xi_1} \frac{\partial \xi_1}{\partial x'^{\mu}}
, \\
\Xi'^{(2)}_{\mu}(x,x') &= \frac{|y'|}{\xi_2} \frac{\partial \xi_2}{\partial x'^{\mu}} 
.
\end{aligned}
\end{equation}
In view of the separation of variables into $r$ and $\theta$, we find it convenient to replace $\Xi^{(1)}$ and $\Xi'^{(1)}$ with 
\be
{\mathcal X}_\mu = |y| \frac{\partial \chi}{\partial x^\mu} \ , \; \; \; {\mathcal X}'_\mu = |y'| \frac{\partial \chi}{\partial x'^\mu} \ ,
\ee
such that ${\mathcal X} = 2 \xi_1 \Xi^{(1)} +  \xi_2 \Xi^{(2)}$ and similarly for ${\mathcal X}'.$\footnote{%
Note ${\mathcal X}$ was used to indicate a different tensor structure in \cite{Herzog:2020bqw}.
} 

The last tensor structure we need is a bivector, constructed by taking both an $x$ and an $x'$ derivative of the invariant cross ratios.
The double derivative of $\xi_2$ can be reconstructed\footnote{This is true when $q=2$, see \cite{Herzog:2020bqw}.} from a product of $\Xi^{(2)}$ and $\Xi'^{(2)}$ and does not give us anything new.
The double derivative of $\xi_1$ on the other hand is the inversion tensor:
\begin{subequations} \label{bivectors}
	\begin{align}
   I_{\mu \nu}(x-x') &= -2\xi_1|y| \frac{\partial}{\partial x^{\mu}} \Xi'^{(1)}_{\nu}=-2\xi_1|y'| \frac{\partial}{\partial x'^{\mu}} \Xi^{(1)}_{\nu}, 
   	\end{align}
\end{subequations}
where $I_{\mu \nu}(x) = \delta_{\mu \nu} - \frac{2x_{\mu}x_{\nu}}{x^2}$.  This tensor plays an important role in conformal field theories
without boundaries and defects.  
As it happens, we find it simpler in what follows to use the double derivative of $\chi$ instead
\be
{\mathbb I}_{\mu\nu} = |y'| \frac{\partial}{\partial x'^\nu} {\mathcal X}_\mu  = |y|  \frac{\partial}{\partial x^\mu} {\mathcal X}'_\nu  \ ,
\ee
even though ${\mathbb I}_{\mu\nu}$ is a more complicated object than $I_{\mu\nu}$.  For one, it is not symmetric in its indices.
Note that 
\[
{\mathbb I}_{\mu\nu} = -I_{\mu\nu} + 2 \xi_1 \Xi^{(1)}_\mu \Xi'^{(1)}_\nu  - \frac{\xi_2^3}{1-\xi_2^2}\Xi^{(2)}_\mu \Xi'^{(2)}_\nu \ .
\]
With these structures in hand, the general form of the Maxwell field strength two-point function, consistent with defect conformal symmetry, can be written as a sum of parity even and parity odd terms\footnote{The parity odd structures can only appear when the parity symmetry is broken since the two-point function between identical operators is parity even. Note, the factor of $2$ has been added with the $\epsilon$ terms in anticipation of complex coordinates we use throughout the paper.} 
\begin{equation} \label{generalFF}
\begin{split}
\langle &F_{\mu \nu}(x) F_{\alpha \beta}(x') \rangle = \frac{1}{|y|^2 |y'|^2} \bigg[2 g_1 {\mathbb I}_{\mu [\alpha}{\mathbb I}_{\beta] \nu} + 4 g_2 {\mathcal X}_{[\nu} {\mathbb I}^{\phantom{(1)}}_{\mu] [\alpha} {\mathcal X}_{\beta]} \\
+ & 4 g_3 \left( {\mathcal X}_{[\nu} {\mathbb I}^{\phantom{(1)}}_{\mu] [\alpha} \Xi^{(2)}_{\beta]} + \Xi^{(2)}_{[\nu} {\mathbb I}^{\phantom{(1)}}_{\mu] [\alpha} {\mathcal X}_{\beta]} \right) + 4 g_4 \Xi^{(2)}_{[\nu} {\mathbb I}^{\phantom{(2)}}_{\mu] [\alpha} \Xi'^{(2)}_{\beta]} +4 g_5 {\mathcal X}^{(1)}_{[\mu}\Xi^{(2)}_{\nu]} {\mathcal X}'_{[\alpha} \Xi'^{(2)}_{\beta]} \bigg] \\
&  + \frac{2}{|y|^2 |y'|^2}  {\epsilon_{\alpha \beta}}^{\gamma \delta} \bigg[2 \tilde g_1 {\mathbb I}_{\mu [\gamma}{\mathbb I}_{\delta] \nu} + 4 \tilde g_2 {\mathcal X}_{[\nu} {\mathbb I}^{\phantom{(1)}}_{\mu] [\gamma} {\mathcal X}_{\delta]} \\
+ & 4 \tilde g_3 \left( {\mathcal X}_{[\nu} {\mathbb I}^{\phantom{(1)}}_{\mu] [\gamma} \Xi^{(2)}_{\delta]} + \Xi^{(2)}_{[\nu} {\mathbb I}^{\phantom{(1)}}_{\mu] [\gamma} {\mathcal X}_{\delta]} \right) + 4 \tilde g_4 \Xi^{(2)}_{[\nu} {\mathbb I}^{\phantom{(2)}}_{\mu] [\gamma} \Xi'^{(2)}_{\delta]} +4 \tilde g_5 {\mathcal X}^{(1)}_{[\mu}\Xi^{(2)}_{\nu]} {\mathcal X}'_{[\gamma} \Xi'^{(2)}_{\delta]} \bigg] \ .
\end{split}
\end{equation} 
Additionally, the parity odd terms are further constrained by Bose symmetry associated with identical operators appearing in the two-point function,
\begin{equation}\label{parityoddBose}
\tilde{g}_3 = 0 \ , \qquad
 \tilde{g}_4 = \cot^2 \theta  \left(\chi \tilde{g}_1 + (\chi^2 -1) \tilde{g}_2 \right) \ , \qquad 
 \tilde{g}_5 = - \cot^2 \theta \left(\tilde{g}_1 + \chi \tilde{g}_2 \right) \ ,
\end{equation}
and hence there are only two independent parity odd structures. 

Assuming that $F_{\mu\nu}$ is free in the bulk, we can apply Maxwell's equations to the two-point function:
\begin{equation}
\partial_\mu \langle F^{\mu\nu}(x) F^{\lambda \rho}(x') \rangle = 0  \; , \; \; \;
\partial_\mu \langle \tilde F^{\mu\nu}(x) \tilde F^{\lambda \rho}(x') \rangle = 0  \ ,
\end{equation}
where $\tilde F^{\mu\nu}(x) = \frac{1}{2} \epsilon^{\mu\nu\lambda \rho} F_{\lambda \rho}(x)$.
By contracting the result with the tensor structures $\Xi^{(2)}$, ${\mathcal X}$, and ${\mathbb I}_{\mu\nu}$ along with the epsilon symbol to make scalar quantities, 
we can construct a set of
coupled linear partial differential equations for the functions $g_i$ and $\tilde g_i$ in the cross ratios $\xi_2$ and $\chi$.  
We find eight coupled equations for the $g_i$ and another independent but identical set of equations for the $\tilde g_i$.\footnote{%
	$\tilde{g}_i$'s are further constrained by \eqref{parityoddBose}. 
}  
Focusing on the $g_i$, 
the four equations $\partial_\mu \langle \tilde F^{\mu\nu}(x) \tilde F^{\lambda \rho}(x') \rangle = 0$ allow us to solve for $g_2$, $g_3$, and $g_5$ in terms of $g_1$ and $g_4$.  The remaining four equations $\partial_\mu \langle  F^{\mu\nu}(x)  F^{\lambda \rho}(x') \rangle = 0$ allow us to solve for $g_1$ and $g_4$.  
The result is that 
\begin{eqnarray}
\label{g2g3g5}
g_2 &=& \frac{1}{2} \frac{\partial g_1}{\partial \chi} \ , \; \; \; g_3 = \frac{\xi_2}{2} \frac{\partial g_1}{\partial \xi_2} \ , \\
g_4 &=& \frac{\cot^2 \theta}{2} \left( -\frac{\partial}{\partial \chi} (\chi^2-1) g_1 +H(\chi) \right)
 \ , \nonumber \\
g_5 &=& \frac{\cot^2 \theta}{2}\left( \frac{1}{\chi} \frac{\partial}{\partial \chi} \chi^2 g_1 + H'(\chi) \right)
\ , \nonumber
\end{eqnarray}
where $H(\chi)$ and $g_1(\chi, \theta)$ satisfy the differential equations
\be
\label{Heq}
(1-\chi^2) H''(\chi) - 5 \chi H'(\chi) - 3 H(\chi) &=& 0 \ , \\
\label{geq}
\left[(1-\chi^2) \partial_\chi^2 - \partial_{\theta}^2  - 5 \chi \partial_\chi  - 4 \right] g_1(\chi, \theta) &=& 0 \ .
\ee
We note in passing that in the parity odd sector, the tilde'ed version of this set of equations along with (\ref{parityoddBose})
restricts $\tilde g_1(\chi, \theta)$ to be a function of $\chi$ only and further sets $\tilde H = 2 \partial_\chi [ (\chi^2 - 1) \tilde g_1 ] $.    

The two differential equations (\ref{Heq}) and (\ref{geq}) are straightforward to solve.  Let us start with the partial differential equation for $g_1(\chi, \theta)$, which succumbs to a separation of variables approach.  Setting $g_1\sim e^{i s \theta} $ and $s \neq 0$ or 1, we find
\be
g_1 =  e^{i s \theta} \frac{r^2 }{(1-r^2)^3} \left( c_- r^{s} (1+s + r^2(1-s)) + c_+ r^{-s} (1-s + r^2(1+s)) \right) \ .
\ee
In the two special cases $s=0$ and 1, the two independent solutions become degenerate, and we find instead extra logarithmic solutions.
For $s=0$, we find
\be
g_1 = \frac{r^2 }{(1-r^2)^3} \left( c (1+r^2)  + c' (2 + (1+r^2) \log r) \right) \ ,
\ee
while for $s=1$, we find instead
\be
g_1 =  e^{i \theta} \frac{r^2 }{(1-r^2)^3} \left( c r + \frac{c'}{r}  (-1 + r^4 - 4 r^2 \log r) \right)\ .
\ee
Intriguingly, $H$ satisfies the same differential equation as $g_1$ in the cases $s= \pm 1$.  As we are dealing with a conformal field theory without extra scales, 
we should set the logarithmic solutions to zero, i.e.\ $c'=0$.  

Moving forward, let us define the mode function 
\be
G_s(r) = \frac{r^{2-s} }{(1-r^2)^3} (1-s + r^2(1+s)) \ .
\ee
The most general solution for $g_1$ can thus be written
\be
\label{gform}
g_1 = c_0 G_0(r) + (c_1 e^{i \theta} + c_{-1} e^{-i \theta}) G_1(r)  + \sum_{s \neq 0,1} e^{i s \theta} (c_{s+} G_{|s|}(r) + c_{s-} G_{-|s|}(r) ) \ .
\ee
The remaining $g_i$ can then be reconstructed from the relations (\ref{g2g3g5}), where we must also make a choice of integration constant
$c_h$ in the solution for 
$H = c_h G_1(r)$ that appears
in the expressions for $g_4$ and $g_5$.  
If we insist on periodicity under $\theta \to \theta + 2 \pi$, then we can restrict the sum to integer $s$.\footnote{%
More generally, we may have a monodromy
type defect with a phase factor $e^{2 \pi i \beta}$ associated with going around the circle.  In this case, we could restrict to 
$s \in {\mathbb Z} + \beta$.
}  

The scaling behavior of the mode function $G_s(r)$ in the limits $r \to 0$, $r \to \infty$, and $r \to 1$ has physical meaning.
In the limit $r\to 0$, the point $x$ is much closer to the defect than $x'$.  In contrast, in the limit $r \to \infty$, the point $x$ is much further away.  In the first instance, we can replace $F_{\mu\nu}(x)$ with a defect operator product expansion (defect OPE), while in the second instance, $F_{\alpha \beta}(x')$ can be replaced.  In the limit $r \to 1$, $x$ and $x'$ become approximately coincident, and the mode functions diverge.  Because of the singularity at $r=1$, we can choose 
different mode functions in the two regions, $0<r<1$ and $r>1$, independently adapted to the boundary conditions
in the limits $r\to 0$ and $r \to \infty$.\footnote{%
\label{fn:Gchi}
The mode function can be given a more uniform description in the $\chi$ variable:
\[
G_s(\chi) =-\frac{1}{4} (\chi + \sqrt{\chi^2-1})^{-s} (\chi + s \sqrt{\chi^2-1}) (\chi^2 - 1)^{-3/2} \ .
\]
Because of the branch cut at $\chi = 1$, the mode function will simplify alternately to $G_s(r)$ or $-G_{-s}(r)$ depending on whether $r>1$ or $0<r<1$.  
}

The point of the next two sections is to elucidate the defect OPE in much greater detail.  At this point, let us instead state the result.  Note that $G_{|s|}(r)$ is well behaved in the limit $r \to \infty$ while $G_{-|s|}(r)$ is well behaved in the limit $r \to 0$.  
There is a two part claim, one for $s \neq 0$ and one for $s = 0$.  For $s \neq 0$, in the region $r>1$, $G_{|s|}(r)$ corresponds to the contribution of a defect vector of dimension $\Delta = 1 + |s|$ while in the region $0<r<1$, $G_{-|s|}(r)$ represents the same contribution.  
For $s=0$, $G_0(r)$, $H(r)$, and $\tilde g_1(r)$ correspond to the contributions of defect scalars of dimension 2.

This two point function building block $g_1$ (\ref{gform}) is similar in structure to the full two-point function of a massless 
free scalar in the presence of a defect.
It was stressed in ref.\ \cite{Lauria:2020emq} that for a free scalar field, the existence of a nontrivial defect
was closely correlated with the ability to have both $c_{s+}$ and $c_{s-}$ simultaneously be nonzero, a freedom we do not have
here except possibly in the $s=0$ case.  
From both the small $r$ and large $r$ scaling of $G_s(r)$,
we can deduce the existence of a defect vector with dimension $\Delta = 1 \pm s$.  Since the unitarity
bound for such an operator is one, for unitary defects we must set either $c_{s+}$ or $c_{s-}$ to zero, depending on which
region we are in, $r>1$ or $0<r<1$.
If we are to extrapolate the lessons of \cite{Lauria:2020emq}, the dimension two scalars associated with $s=0$
would seem to provide the only flexibility to have a nontrivial defect and will need careful consideration in what follows.
First though we would like to discuss 
the one-point function of the stress-tensor.

\subsection{Stress Tensor}

From these differential equations, we can deduce that 
\be
\langle F_{\mu\nu}(x) F^{\mu\nu}(x') \rangle &=& \frac{1}{ |y|^2 |y'|^2} \biggl(
(\chi + 2 \xi_2) H + (\chi^2-1) H'  + \nonumber \\
&& \hspace{1in}  - \frac{({\bf x} - {\bf x}')^2}{|y| |y'|} (H + (\chi - \xi_2) H') \biggr)
\ee
The stress tensor $T_{\mu\nu} = F_{\mu\rho} {F_{\nu}}^\rho - \frac{1}{4} \delta_{\mu\nu} F_{\lambda \rho} F^{\lambda \rho}$ can be constructed
from the coincident limit of the Green's function.  We find, evaluating $g_1(\chi, \xi_2)$ and $H(\chi)$ at specific points,
\be
\langle T_{\mu\nu}(x) \rangle  = \frac{g_1(1,1)-\frac{1}{4} H( 1) }{y^4} \left(-4\mathcal{J}_{\mu\nu} + \delta_{\mu\nu} \right) \ ,
\ee
where ${\mathcal J}_{\mu\nu} \sim \Xi^{(2)}_\mu \Xi^{(2)}_\nu$ is the bitensor 
\be
{\mathcal J}_{\mu\nu} = \begin{cases}
\delta_{ij} - n_i n_j & \mu=i, \nu=j \\
0 & \mbox{otherwise}
\end{cases}
\ee
where $i$ and $j$ index directions normal to the defect and $n_\mu \equiv y_\mu / |y|$.
The $\tilde g_i$ do not contribute to $\langle F_{\mu\nu}(x) F^{\mu\nu}(x') \rangle$ nor to $\langle T_{\mu\nu}(x) \rangle$.

The functions $G_s(r)$, into which $g_1(\chi, \theta)$ and $H(\chi)$ can be decomposed, are divergent at $\chi=1$ (equivalently $r=1$) and so we need to be careful with this expression for 
the stress tensor expectation value.  The simplest theory -- Maxwell theory in the absence of a defect -- has 
 \cite{Herzog:2020bqw}
\be
g_1(\chi, \xi_2) = c \, \xi_1^{-2} = c \frac{16}{(\chi -\xi_2)^2}= c \left( \frac{4r}{1 + r^2 - 2 r \cos(\theta)} \right)^2\ , 
\ee
where we set $c=1$ in what follows.   (Note also $H = 0$ and $\tilde g_1 = 0$.)
Clearly $g_1(1,1)$ is divergent, while based on Lorentz invariance, we expect the stress-tensor
expectation value to vanish. 

The situation suggests a minimal subtraction prescription.  Whenever we 
compute $\langle T_{\mu\nu} \rangle$, we regulate by subtracting the 
$\langle T_{\mu\nu} \rangle$ of the no defect theory.
Note that the no defect $g_1$ has the following mode decomposition:
\be
\xi_1^{-2} = -16  \sum_{j \in {\mathbb Z}}  e^{ i j \theta} G_{|j|} \ .
\ee
Decomposing $g_1(\chi, \xi_2)$ into a mode sum, one possible generalization of the no defect theory
with a finite stress tensor involves introducing $H$ and keeping $c_{1} + c_{-1} -\frac{1}{4}c_h = -32$.  

In the next subsection, we consider some simple orbifold theories to understand how this minimal
subtraction works in an example.

\subsection{Orbifold Theories} \label{orbifold}

For a ${\mathbb Z}_m$ orbifold theory, we can obtain the Green's function as a sum over no defect theory images
(see  \cite{Herzog:2020bqw}), where we not only have an insertion of
$F_{\mu\nu}(x)$ at $\theta$ but also at $\theta + \frac{2\pi k}{m}$ for $k = 1, 2, \ldots, m-1$.  For each of these images, we have a contribution
to the Green's function from a $g_1$ of the form
\be
g_1^{(\alpha)} = \left( \frac{4 r}{1+r^2 - 2r \cos (\theta + \alpha)} \right)^2
\ee
where $\alpha = \frac{2\pi k}{m}$.  
Each such $g_1^{(\alpha)}$ admits a mode decomposition 
\be
g_1^{(\alpha)} = -16 \sum_{j \in {\mathbb Z}}  e^{ i j (\theta +  \alpha)} G_{|j|} \ .
\ee
The full orbifold theory is then a sum over the appropriate $g_1^{(\alpha)}$:
\be
g_1^{\langle m \rangle} = \sum_{k=0}^{m-1} g_1^{\left( \frac{2 \pi k}{m} \right)} = -16 \sum_{s \in {\mathbb Z}} e^{i s \left( \pi - \frac{\pi}{m} + \theta \right)}
\frac{\sin \pi s}{\sin \left(\frac{\pi s}{m} \right)} G_{|s|} = -16 m \sum_{s \in m{\mathbb Z}}  e^{i s \theta} G_{|s|}  \ ,
\ee
where all the $s$ not divisible by $m$ get removed from the sum, making the function periodic under $\theta \to \theta + \frac{2 \pi}{m}$.  

Returning to the issue of the finiteness of the stress tensor, we see that $g_1^{(\alpha)}$ will in general be finite in the coincident limit
 $\theta = 0$ and $r=1$.  
The one exception is the defect free case, 
where $\alpha = 0$.  
As discussed above,  we can regulate
the orbifold result by subtracting the no defect result.  The stress tensor expectation value is then obtained by summing over all the $\alpha \neq 0$.  
The result for the ${\mathbb Z}_m$ orbifold theory is
\be
\langle T_{\mu\nu} \rangle = \frac{(m^2+11)(m+1)(m-1)}{45 y^4}  \left( -4 \mathcal{J}_{\mu\nu} + \delta_{\mu\nu} \right) \ .
\ee

\section{Two Point Functions Involving Defect Operators}
\label{sec:FO}

We show that the only defect operators with which a Maxwell field can have a nonzero two-point function are a scalar of dimension two or an operator with spin
one along the defect.
To obtain correlation functions involving defect operators, we will follow the strategy outlined in \cite{Herzog:2020bqw}, and take boundary limits of the bulk 
tensor structures described above.  In particular, we use the following tensor building blocks (see appendices \ref{sec:U1structs} and \ref{sec:bulkn})
\be
\Xi^{(1)}_\mu \ , \; \; \;
{\mathcal I}_{\mu \nu} \equiv I_{\mu\nu} - \xi_2 \Xi^{(1)}_\mu (\Xi'^{(1)}_\nu - \Xi'^{(2)}_\nu) \ , \; \; \;
\mathcal{J}_{\mu\nu} \ , \; \; \;
\delta_{\mu\nu} \ .
\ee
We then take the boundary limit of the $x'$ insertion, leading to the structures
\be
\hat \Xi^{(1)}_\mu = \lim_{y' \to 0} \Xi^{(1)}_\mu \ , \; \; \;
\hat {\mathcal I}_{\mu \nu} =  \lim_{y' \to 0}  {\mathcal I}_{\mu \nu}  \ ,
\ee
while the tensors $\mathcal{J}_{\mu\nu}$ and $\delta_{\mu\nu}$ are untouched by this limiting procedure.  We claim that this list is sufficient to construct
the two-point functions of interest.  
Finally, we work in complex coordinates, replacing $y_1$ and $y_2$ with $z = y_1 + i y_2$ and $\bar z = y_1 - i y_2$, and similarly for $\textbf{x} \rightarrow (w, \bar w)$.  This change of variable
will make the representation of the boundary fields under the transverse and parallel $SO(2)$ rotation groups plain.  Indeed, it is sufficient to include a 
factor of $(z / |z|)^s$ to give an operator spin $s$ in the transverse directions. 
Table \ref{bulk-to-defect} summarizes the transformation properties of these tensor building blocks.
\begin{table}[]
	\centering
	\begin{tabular}{|cccc|}
		\hline
		\hline
		\hline
		\multicolumn{4}{|c|}{\textbf{Bulk-Defect $U(1)$ Structures}}                                                                                                                                                                                                                                                \\ \hline \hline \hline
		\multicolumn{3}{|c|}{At $x$}                                                                                                                                                                                                                                       & At $\textbf{x}'$                       \\ \hline 
		\multicolumn{1}{|c|}{$\ell=s=0$}                                                                                                                       & \multicolumn{1}{c|}{$\ell=-1$}                    & \multicolumn{1}{c|}{$\ell=1$}                         & $\ell=0, s$                            \\ \hline
		\multicolumn{1}{|c|}{$\hat{\Xi}^{(1)}_{\mu}, \frac{\bar z}{|z|} \hat{\mathcal{I}}_{\mu \bar z} ,\hat{\mathcal{I}}_{[\mu| w} \hat{\mathcal{I}}_{|\nu] \bar{w}} $} & \multicolumn{1}{c|}{$\hat{\mathcal{I}}_{\mu w}$} & \multicolumn{1}{c|}{$\hat{\mathcal{I}}_{\mu \bar{w}}$} & $\left(\frac{z}{|z|}  \right)^s$ \\ \hline \hline \hline
	\end{tabular}
	\caption{The tensor structures necessary for building bulk-defect two-point functions, along with their transformation properties with respect to the
	parallel and transverse $SO(2)$ rotation groups; $\ell$ is used for the parallel directions and $s$ for the transverse ones. 
	The bulk indices are $\mu$ and $\nu$.  The defect indices are $w$ and $\bar w$ along the defect and $z$ and $\bar z$ transverse to it. More details on the tensor structures can be found in \eqref{bulkdefectU1}.}
\label{bulk-to-defect}
\end{table}

\subsection{$\left \langle F \hat{\mathcal{O}}^{(s)}_{( \Delta)} \right \rangle$}
We consider a defect scalar $\hat{\mathcal O}^{(s)}_{( \Delta)}$ with transverse spin $s$ and dimension
$ \Delta$.  
\begin{equation} \label{FhatOs}
\begin{aligned}
\langle F_{\mu \nu}(x) \hat{\mathcal{O}}^{(s)}_{(\hat \Delta)}(\textbf{x}') \rangle = \frac{2}{|x-\textbf{x}'|^{2{\Delta}} |z|^{2 - {\Delta}}} \bigg[&c_{F {\mathcal{O}}} \hat{\mathcal{I}}_{[\mu| w} \hat{\mathcal{I}}_{| \nu] \bar w} \left(\frac{z}{|z|} \right)^s \\
&+ c'_{F {\mathcal{O}}} \hat{\Xi}^{(1)}_{[\mu} \hat{\mathcal{I}}_{\nu] \bar z} \left(\frac{z}{|z|} \right)^{s-1} \bigg],
\end{aligned} 
\end{equation}
where $|x-\textbf{x}'|^2 = |w-w'|^2+|z|^2$. Applying the equation of motion gives the constraints $(-2+{\Delta}+s)c'_{F {\mathcal{O}}}=0$, $(-2+{\Delta}-s)c'_{F {\mathcal{O}}}=0$ and $s c'_{F  {\mathcal{O}}} = 0$. Further applying the Bianchi identity gives the constraints $(-2+{\Delta}+s)c_{F {\mathcal{O}}}=0$, $(-2+{\Delta}-s)c_{F {\mathcal{O}}}=0$ and $s c_{F {\mathcal{O}}} = 0$. The simplest solution is $s=0$ and ${\Delta}=2$. We could also solve all the equation of motion constraints by setting $c'_{F {\mathcal{O}}}=0$. However now solving the Bianchi identity constraints requires either $c_{F {\mathcal{O}}}=0$ where the correlator vanishes completely or $s=0$ and ${\Delta}=2$, i.e.~the solution we had mentioned first. So in general a free Maxwell field only has non-zero correlator with a defect scalar with dimension ${\Delta}=2$ and $s=0$.
\subsection{$\left \langle F \hat{\mathcal{W}}^{(s)}_{(\Delta, \ell)}\right \rangle $}
\label{sec:FW}
Next we consider a defect vector $\hat {\mathcal W}^{(s)}_{(\Delta, \ell)} $ with parallel spin $\ell=1$, dimension $\Delta$, 
and transverse spin $s$:
\begin{equation}
\begin{aligned}
\left \langle F_{\mu\nu}(x)\hat{\mathcal{W}}^{(s)}_{(\Delta, 1)}(\textbf{x}') \right \rangle =  \frac{2}{|x-\textbf{x}'|^{2{\Delta}} |z|^{2- {\Delta}}} \bigg[c_{F {\mathcal{W}}} \hat{\Xi}^{(1)}_{[\mu} \hat{\mathcal{I}}_{\nu] \bar w} \left(\frac{z}{|z|} \right)^s \\
+ c'_{F {\mathcal{W}}} \hat{\mathcal{I}}_{[\mu| \bar w} \hat{\mathcal{I}}_{|\nu] \bar z} \left(\frac{z}{|z|} \right)^{s-1} \bigg].
\end{aligned}
\end{equation}
Applying the equation of motion gives the constraint $2(-1+{\Delta})c_{F {\mathcal{W}}} -  sc'_{F {\mathcal{W}}} = 0$. Further applying the Bianchi identity gives the constraint $(-1+{\Delta})c'_{F {\mathcal{W}}} - 2s c_{F {\mathcal{W}}} = 0$. These coupled equations can be solved to give the following two sets of solutions:
\begin{equation} \label{constraintofdimensionFW}
\{ {\Delta} = 1 - s, \quad c'_{F {\mathcal{W}}} = -2  c_{F {\mathcal{W}}} \}, \quad \{ {\Delta} = 1 + s, \quad c'_{F {\mathcal{W}}} = 2  c_{F {\mathcal{W}}} \}. 
\end{equation}
These solutions indicate two series of parallel spin $\ell=1$ and transverse spin $s$ defect primaries $\hat{\psi}_{\pm}^{(s)}$ with dimension ${\Delta}_{\pm}=1\pm s$. Reflection positivity requires that ${\Delta} \geq p+|\ell|-2=1$ which gives us the constraints,
\begin{equation}
\begin{cases}
\hat \psi_+^{(s)} \implies s\geq 0,  \\
\hat \psi_-^{(s)} \implies s \leq 0.
\end{cases}
\end{equation}
It is convenient to separate out the spin $s=0$ case and combine the remaining two sets of operators: $\hat{\psi}_+^{(s)}$ and $\hat{\psi}_-^{(-s)}$ for $s>0$ with dimension ${\Delta}=1+s$, alongside $\hat{\psi}_+^{(0)}$ and $\hat{\psi}_-^{(0)}$ with dimension $1$. The non-zero defect two-point function is then $\langle \hat \psi_+^{(s)} \hat \psi_-^{(-s)} \rangle$ for $s>0$. While any of the $s=0$ vector modes can have a non-zero two-point function, we will show in section \ref{sec:conformalblock} that they decouple from the defect OPE of $F_{\mu \nu}$ and only the scalar with $\Delta = 2$ contributes when $s=0$. 

The correlator with a spin $\ell=-1$ vector is given by,
\begin{equation}
\begin{aligned}
\left \langle F_{\mu\nu}(x)\hat{\mathcal{W}}_{(\Delta, -1)}^{(s)}(\textbf{x}') \right \rangle =  \frac{2}{|x-\textbf{x}'|^{{2\Delta}} |z|^{2- {\Delta}}} \bigg[c_{F {\mathcal{W}}} \hat{\Xi}^{(1)}_{[\mu} \hat{\mathcal{I}}_{\nu] w} \left(\frac{z}{|z|} \right)^s \\
+ c'_{F {\mathcal{W}}} \hat{\mathcal{I}}_{[\mu| w} \hat{\mathcal{I}}_{|\nu] \bar z} \left(\frac{z}{|z|} \right)^{s-1} \bigg] \ .
\end{aligned}
\end{equation}
Applying the equation of motion and Bianchi identity gives the exact same solution as \eqref{constraintofdimensionFW} and hence there again exists two sets of operators: $\hat{\phi}_+^{(s)}$ and $\hat{\phi}_-^{(-s)}$ for $s > 0$ with spin $\ell = -1$ and ${\Delta} = 1+s$, alongside $\hat{\phi}_{\pm}^{(0)}$. 
An important relation is the identification
$\hat{\phi}_{\mp}^{(\mp s)} = (\hat{\psi}_{\pm}^{(\pm s)})^*$, where $^*$ is complex conjugation. 

Lastly, the correlation functions $\left \langle F_{\mu\nu}(x)\hat{\mathcal{W}}_{(\Delta, \ell)}^{(s)}(\textbf{x}') \right \rangle$  with parallel spin operators $|\ell| \geq 2$ vanish. 
\subsection{Parity Odd Structures}
We claim our bulk-defect tensor structures are the most general possible and allow us to reproduce correlation
functions of arbitrary parity. 
We expect contracting parity even tensor structures with $\epsilon_{\mu \nu \rho \sigma}$ will generate parity odd structures, and vice versa. 
Contracting $\epsilon_{\mu \nu \rho \sigma}$ with our bulk-defect tensor structures 
does not generate new structures, however.  Instead, the contraction 
interchanges the structures we have.  
As shown in section \ref{sec:conformalblock}, the defect OPE obtained using these give rise to both the parity odd and parity even conformal blocks contributing to $\langle F_{\mu \nu}(x) F_{\alpha \beta}(x')  \rangle$. 
\subsection{Defect Two-Point Function}

If we have two defect operators $\hat {\mathcal W}^{(s)}_{(\Delta, \ell)}(\textbf{x})$ and $\hat {\mathcal W}^{(s')}_{(\Delta', \ell')}(\textbf{x}')$, the conformal Ward identities constrain the two point function to have the form
\be
\langle \hat {\mathcal W}^{(s)}_{(\Delta, \ell)}(\textbf{x})\hat {\mathcal W}^{(s')}_{(\Delta', \ell')}(\textbf{x}') \rangle = c \frac{\delta_{\ell, \ell'} \delta_{s, -s'} \delta_{ \Delta,  \Delta'}}{(w-w')^{ \Delta - \ell} (\bar w - \bar w')^{ \Delta + \ell}} \ , 
\ee
where $c$ is an over-all normalization constant.
The $s = 0 =s'$ case is familiar from two dimensional conformal field theory.  The extra information here is that the two-point function will vanish unless $s+s' = 0$.

\section{Defect OPE and Conformal Blocks}
\label{sec:conformalblock}

Here, we use the bulk-defect two-point functions from section \ref{sec:FO} to deduce the defect OPE
of $F_{\mu\nu}$.  The defect OPE then allows us to recognize the mode decomposition of
 $\langle F_{\mu \nu}(x) F_{\alpha \beta}(x')  \rangle$ that we had earlier as a decomposition 
into conformal blocks.  
The defect OPE will also be crucial information for our calculation of the bulk-defect-defect three-point function
in section \ref{sec:FWW}.

\subsection{Defect OPE: Vectors}

Let us begin by considering the contribution of $\ell=\pm1$ defect operators $\hat{\psi}_\pm^{(\pm s)}(\textbf{x})$ and $\hat{\phi}_\pm^{(\pm s)}(\textbf{x})$
discussed in section \ref{sec:FW} 
to the defect OPE of $F_{\mu\nu}$:  
\begin{equation}
\label{FdefectOPE}
\begin{aligned}
F_{\mu \nu}(x) \big{|}_{\ell = \pm 1}= \sum_{s \geq 0 }\bigg( &\mathcal{A}_{\mu \nu}^{(s)}(z, \bar{z}, \partial_{\textbf{x}}) \hat{\psi}_+^{(s)}(\textbf{x})  
+\mathcal{B}_{\mu \nu}^{(s)}(z, \bar{z}, \partial_{\textbf{x}}) \hat{\psi}_-^{(-s)}(\textbf{x})   \\
+ &\mathcal{C}_{\mu \nu}^{(s)}(z, \bar{z}, \partial_{\textbf{x}}) \hat{\phi}_+^{(s)}(\textbf{x})  
+\mathcal{D}_{\mu \nu}^{(s)}(z, \bar{z}, \partial_{\textbf{x}}) \hat{\phi}_-^{(-s)}(\textbf{x}) \bigg) \ .
\end{aligned}
\end{equation}
It will be enough to focus on just a few specific tensorial components of the defect OPE.
The corresponding bulk-defect two point functions for $\ell=1$ are
\begin{equation} \label{bulkdefectFpsi}
\begin{aligned}
&\langle F_{w \bar{w}}(x) \hat{\psi}_+^{(s)}(\textbf{x}') \rangle = \frac{c^{(s)}_{F+}}{2} \frac{z^s(w'-w)}{|x-\textbf{x}'|^{4+2s}}, \quad  \langle F_{z \bar{z}}(x) \hat{\psi}_+^{(s)}(\textbf{x}') \rangle = \frac{c^{(s)}_{F+}}{2}  \frac{z^s(w-w')}{|x-\textbf{x}'|^{4+2s}}, \\
&\langle F_{w \bar{w}}(x) \hat{\psi}_-^{(-s)}(\textbf{x}') \rangle = \frac{c^{(s)}_{F-}}{2} \frac{\bar z^{s}(w'-w)}{|x-\textbf{x}'|^{4+2s}}, \quad \langle F_{z \bar{z}}(x) \hat{\psi}_-^{(-s)}(\textbf{x}') \rangle = \frac{c^{(s)}_{F-}}{2}  \frac{\bar z^{s}(w'-w)}{|x-\textbf{x}'|^{4+2s}}, 
\end{aligned}
\end{equation}
and for $\ell = -1$ are
\begin{equation}
\label{bulkdefectFphi}
\begin{aligned}
&\langle F_{w \bar w}(x) \hat{\phi}_{+}^{(s)}(\textbf{x}') \rangle = \frac{d_{F+}^{(s)}}{2}  \frac{ z^s (\bar w - \bar w')}{|x-\textbf{x}'|^{4+2s}}, \quad \langle F_{z \bar z}(x) \hat{\phi}_{+}^{(s)}(\textbf{x}') \rangle = \frac{d_{F+}^{(s)}}{2}  \frac{ z^s (\bar w - \bar w')}{|x-\textbf{x}'|^{4+2s}}, \\
&\langle F_{w \bar w}(x) \hat{\phi}_{-}^{(-s)}(\textbf{x}') \rangle = \frac{d_{F-}^{(s)}}{2}  \frac{ \bar z^s (\bar w - \bar w')}{|x-\textbf{x}'|^{4+2s}}, \quad \langle F_{z \bar z}(x) \hat{\phi}_{-}^{(-s)}(\textbf{x}') \rangle = - \frac{d_{F-}^{(s)}}{2}  \frac{ \bar z^s (\bar w - \bar w')}{|x-\textbf{x}'|^{4+2s}}.
\end{aligned}
\end{equation}

It can be shown that $d_{F\pm}^{(s)} = (c_{F \mp}^{(s)})^*$ using the fact that $F_{\mu \nu}$ is a real irrep of $SO(4)$ in Cartesian coordinates. For systems with degeneracies, one expects to be able to pair up the $\ell = \pm 1$ operators as a field and its complex conjugate, such that the contribution to the defect OPE is `block diagonalized'. 

Using the bulk-defect two point functions and (\ref{diffops}), we can present a more explicit form for the differential operators in the defect OPE (\ref{FdefectOPE}):
\begin{equation}
\label{ABdefs}
\begin{aligned}
\mathcal{A}_{w \bar{w}}^{(s)} =  {\mathcal A}_{z \bar z}^{(s)} = \frac{\bar z^s}{2}  \sum_{m=0}^{\infty} \frac{(-|z|^2)^m}{m! (s)_{m+1}} \partial_{\bar{w}}^{m} \partial_w^{m+1} \ ,\\
\mathcal{B}_{w \bar{w}}^{(s)} = - \mathcal{B}_{z \bar{z}}^{(s)}= \frac{z^s}{2}  \sum_{m=0}^{\infty} \frac{(-|z|^2)^m}{m! (s)_{m+1}} \partial_{\bar{w}}^{m} \partial_w^{m+1}, 
\end{aligned}
\end{equation} 
and for $\ell =- 1$, 
\begin{equation}
\label{CDdefs}
\begin{aligned} 
\mathcal{C}^{(s)}_{w \bar w} = - \mathcal{C}^{(s)}_{z \bar z}  = - \frac{\bar z^s}{2}  \sum_{m=0}^{\infty} \frac{(-|z|^2)^m}{m!(s)_{m+1}} \partial_{\bar w}^{m+1} \partial_w^m, \\
\mathcal{D}^{(s)}_{w \bar w} = \mathcal{D}^{(s)}_{z \bar z}  = - \frac{z^s}{2}  \sum_{m=0}^{\infty} \frac{(-|z|^2)^m}{m!(s)_{m+1}} \partial_{\bar w}^{m+1} \partial_w^m.
\end{aligned}
\end{equation}
Note we have used the freedom to rescale the fields to fix the normalization of these differential operators.
As a result, the defect two-point function may have a non-canonical normalization:
\be
\label{psipsi}
\langle \hat{\psi}_+^{(s)}(\textbf{x}) \hat{\psi}_-^{(-s)} (\textbf{x}')  \rangle = c^{(s)}_{+-} \frac{(w-w')^2}{|\textbf{x}-\textbf{x}'|^{2s+4}} \ ,
\ee
and the complex conjugate for $\phi_{\pm}^{(\pm s)}$.
Consistency of (\ref{bulkdefectFpsi}), (\ref{bulkdefectFphi}), (\ref{ABdefs}), (\ref{CDdefs}) and (\ref{psipsi}) 
then requires 
\begin{equation}
c^{(s)}_{F+} = c^{(s)}_{F-} = c^{(s)}_{+-}, \qquad d^{(s)}_{F+} = d^{(s)}_{F-} = \bar{c}^{(s)}_{+-},
\end{equation}
where $\bar{c}^{(s)}_{+-}$ is the complex conjugate of $c^{(s)}_{+-}$. 

Certain components of these bulk-defect two-point functions $\langle F_{\mu\nu} \hat \psi^{(s)}_\pm \rangle$ and $\langle F_{\mu\nu} \hat \phi^{(s)}_\pm \rangle$ vanish, which means that certain components of the 
tensorial differential operators ${\mathcal A}_{\mu\nu}^{(s)}$,  ${\mathcal B}_{\mu\nu}^{(s)}$,  ${\mathcal C}_{\mu\nu}^{(s)}$,  and ${\mathcal D}_{\mu\nu}^{(s)}$ must vanish as well.  We will have occasion to make use of these conditions in the next section, when we consider the bulk-defect-defect three-point function
$\langle F_{\mu\nu} \hat {\mathcal W}_1 \hat {\mathcal W}_2 \rangle$.  These conditions are that
\be
\label{vanishingABCD}
{\mathcal A}_{w \bar z}^{(s)} = 0 ={\mathcal A}_{\bar w  z}^{(s)} \ , \; \; \;
{\mathcal B}_{w z}^{(s)} = 0 ={\mathcal B}_{\bar w  \bar z}^{(s)}  \ , \\
{\mathcal C}_{w z}^{(s)} = 0 ={\mathcal C}_{ \bar w  \bar z}^{(s)} \ , \; \; \;
{\mathcal D}_{w \bar z}^{(s)} = 0 ={\mathcal D}_{\bar w  z}^{(s)}  \ . \nonumber
\ee

Note that the $s=0$ defect-defect two-point function  $\langle \hat \psi^{(0)}_+ \hat \psi^{(0)}_- \rangle$ is purely anti-holomorphic and equal to $(\bar w - \bar w')^{-2}$ while the bulk-defect two-point function contains an over-all factor of $((w-w')(\bar w - \bar w') +|z|^2)^{-2}$. It is not possible to act on the $s=0$ defect vector two-point function with 
${\mathcal A}_{\mu\nu}^{(s)}$ or ${\mathcal B}_{\mu\nu}^{(s)}$ and get the corresponding bulk-defect two-point function.
Indeed, because of an extra holomorphic derivative $\partial_w$, ${\mathcal A}_{\mu\nu}^{(s)}$ or ${\mathcal B}_{\mu\nu}^{(s)}$  will annihilate
$\langle \hat \psi^{(0)}_+ \hat \psi^{(0)}_- \rangle$.
In fact ${\mathcal A}_{\mu\nu}^{(s)}$ or ${\mathcal B}_{\mu\nu}^{(s)}$  are not even well-defined at $s=0$ because of a
divergent $1/(s)_{m+1}$ factor.
Similar argument holds for the $\ell = -1$ case. We conclude that the $s=0$ boundary vectors, which are in fact boundary conserved currents, are absent from the boundary OPE of $F_{\mu\nu}$. 
Any $s=0$ contribution to correlation functions of $F_{\mu\nu}$ with other operators 
must come from the $\Delta = 2$ boundary scalars, which we come to next.

\subsection{Defect OPE: Scalars}

It remains to consider the OPE contribution of 
a defect scalar $\hat{\mathcal{O}}^{(s)}_{(\Delta)}$.
As we saw previously, the free field constraints mean $\left \langle F_{\mu\nu} \hat {\mathcal O}^{(s)}_{(\Delta)} \right \rangle = 0$
unless $s = 0$ and $\Delta = 2$. Therefore we expect only defect operators of type
$\hat {\mathcal O}^{(0)}_{(2)}$ to appear in the defect OPE
of $F_{\mu\nu}$.  
Because $s=\ell=0$, it is simplest to work with real scalars. Writing directly in Cartesian coordinates, the two-point function is, 
\begin{equation} \label{FOrealspace}
\begin{aligned}
\langle F_{\mu \nu}(x) \hat{\mathcal{O}}^{(0)}_{(\Delta)}(\textbf{x}') \rangle &= \frac{2}{|x-\textbf{x}'|^{2  \Delta} |y|^{2 -  \Delta}} \left[c_{F \hat{\mathcal{O}}}\hat{\mathcal{I}}_{\mu a} \hat{\mathcal{I}}_{\nu b} \epsilon^{ab} + c'_{F \hat{\mathcal{O}}} \hat{\Xi}^{(1)}_{[\mu|} \hat{\mathcal{I}}_{|\nu] i} \hat{\mathcal{Y}}'_{j} \epsilon^{ij} \right], \\
&= c_{FO} \mathcal{T}^{(1)}_{\mu \nu} + c'_{FO} \mathcal{T}^{(2)}_{\mu \nu} \ ,
\end{aligned}
\end{equation} 
where $\mathcal{Y}'_{\mu} = \xi_2 \left(\Xi^{'(1)}_{\mu} - \Xi^{'(2)}_{\mu} \right)$.\footnote{%
$\mathcal{Y}'$ is the same as $\mathcal{X}'$ in ref.\ \cite{Herzog:2020bqw} and we use $a,b$ for indices on the defect while $i,j$ are transverse to the defect.
} 
Appearance of the $\epsilon$'s makes it clear that $\mathcal{T}^{(1)}_{\mu \nu}$ and $\mathcal{T}^{(2)}_{\mu \nu}$ are parity odd structures with respect to parity parallel and transverse to the defect, respectively. Using this we postulate the existence of two pseudo-scalars, $\parscalar$ and $\perpscalar$, associated with parity parallel and transverse to the defect respectively.

More precisely, we define $\parscalar$ and $\perpscalar$ to be the defect scalars that appear in the defect OPE of $F_{\mu\nu}$ in the fashion
\begin{equation}
\label{Fs0OPE}
F_{\mu \nu}(x)|_{s=0} = \mathcal{D}^{(\parscalarnohat)}_{\mu \nu}(z, \bar z, \partial_{\textbf{x}}) \parscalar(\textbf{x}) + \mathcal{D}^{(\perpscalarnohat)}_{\mu \nu}(z, \bar z, \partial_{\textbf{x}}) \perpscalar(\textbf{x}) \ ,
\end{equation}
where $\mathcal{D}^{(\parscalarnohat )}_{\mu \nu}$ and $\mathcal{D}^{(\perpscalarnohat )}_{\mu \nu}$ acting on $|w-w'|^{-4}$ produce $\mathcal{T}^{(1)}_{\mu \nu}$ and $\mathcal{T}^{(2)}_{\mu \nu}$ respectively. 
We will be more specific about the form of the $\mathcal{D}^{(\parscalarnohat )}_{\mu \nu}$ and $\mathcal{D}^{(\perpscalarnohat )}_{\mu \nu}$ presently.

Now because we have defined $\parscalar$ and $\perpscalar$ via (\ref{Fs0OPE}), we are not guaranteed that they
are canonically normalized or even orthogonal.  In general we will have
\begin{equation}
\begin{aligned}
\langle \parscalar(\textbf{x}) \parscalar(\textbf{x}') \rangle = \frac{c_{\parscalarnohat \parscalarnohat}}{|\textbf{x}-\textbf{x}'|^4}, \quad \langle \parscalar(\textbf{x}) \perpscalar(\textbf{x}') \rangle = \frac{c_{\parscalarnohat \perpscalarnohat}}{|\textbf{x}-\textbf{x}'|^4}, \quad \langle \perpscalar(\textbf{x}) \perpscalar(\textbf{x}') \rangle = \frac{c_{\perpscalarnohat \perpscalarnohat}}{|\textbf{x}-\textbf{x}'|^4} \ .
\end{aligned}
\end{equation}
In the special case of a parity preserving theory, indeed symmetry sets $c_{\parscalarnohat \perpscalarnohat} = 0$.

Returning to the $\mathcal{D}^{(\parscalarnohat )}_{\mu \nu}$ and $\mathcal{D}^{(\perpscalarnohat )}_{\mu \nu}$, they are constructed to reproduce the form of the bulk-defect two-point function.
In components, we have that\footnote{We go back into complex coordinates to simplify calculations.}
\begin{equation}
\label{FhatO}
\begin{aligned}
\langle F_{w \bar{w}}(\omega)\hat  {\mathcal O}^{(0)}_{(2)}(\textbf{x}') \rangle &= -\frac{ic_{F {\mathcal O}}}{2} \left(\frac{|w-w'|^2 - |z|^2}{|x-x'|^6}\right) \ ,\\
\langle F_{z \bar{z}}(\omega) \hat {\mathcal O}^{(0)}_{(2)}(\textbf{x}') \rangle &=  - \frac{ic'_{F {\mathcal O}}}{2} \left(\frac{|w-w'|^2-|z|^2}{|x-x'|^6} \right) \ .
\end{aligned}
\end{equation} 
Thus, we set
\begin{eqnarray}
\label{Drels}
\mathcal{D}^{(\parscalarnohat)}_{w \bar w} = -i \frac{\mathcal{D}}{2}, \quad \mathcal{D}^{(\parscalarnohat)}_{z \bar z} = 0, \qquad \mathcal{D}^{(\perpscalarnohat)}_{w \bar w} = 0, \quad \mathcal{D}^{(\perpscalarnohat)}_{z \bar z} = -i \frac{\mathcal{D}}{2}.
\end{eqnarray}
The operator ${\mathcal D}$ is then defined such that
\begin{equation}
\label{Ddef}
\begin{aligned}
\frac{|w-w'|^2-|z|^2}{|x-x'|^6} = \sum_{m=0}^{\infty} \frac{(-|z|^2)^m}{(m!)^2} \partial^m_{\bar{w}} \partial^m_w \left[\frac{1}{|w - w'|^4} \right] \equiv \mathcal{D}\left[\frac{1}{|w - w'|^4} \right] \ .
\end{aligned}
\end{equation}
Consistency of the defect OPE with the bulk-defect two-point functions now requires that
\begin{equation}
c_{F \parscalarnohat} = c_{\parscalarnohat \parscalarnohat}, \quad c'_{F \parscalarnohat} = c_{F \perpscalarnohat} = c_{\parscalarnohat \perpscalarnohat}, \quad c'_{F \perpscalarnohat} = c_{\perpscalarnohat\perpscalarnohat} \ .
\end{equation}

Reflection positivity puts nontrivial constraints on the $2 \times 2$ matrix of $\parscalar$ and $\perpscalar$ two-point functions.  Picking a plane orthogonal to the defect, 
the reflection action will flip the sign of $\parscalar$ but not of $\perpscalar$.   The following matrix must then
have strictly non-negative eigenvalues:
\begin{equation}
\begin{pmatrix}
&-c_{\parscalarnohat \parscalarnohat} &-c_{\parscalarnohat \perpscalarnohat} \\
&c_{\parscalarnohat \perpscalarnohat} &c_{\perpscalarnohat \perpscalarnohat}
\end{pmatrix} \ .
\end{equation}
In other words, the trace  $c_{\perpscalarnohat \perpscalarnohat} - c_{\parscalarnohat \parscalarnohat}  \geq 0$ 
and the determinant $c_{\parscalarnohat \perpscalarnohat}^2  - c_{\parscalarnohat \parscalarnohat}c_{\perpscalarnohat \perpscalarnohat} \geq 0$ must be non-negative.
In the particular case of a parity symmetric theory, where $c_{\parscalarnohat \perpscalarnohat}= 0$, 
we find that the $\parscalar$ two-point function must be negative $c_{\parscalarnohat \parscalarnohat} \leq 0$ and 
the $\perpscalar$ two-point function positive $c_{\perpscalarnohat \perpscalarnohat} \geq 0$.

\subsection{Conformal Blocks}

Here we compute the conformal block contribution of $\hat{\mathcal{O}}^{(0)}_{(\Delta{=}2)}$ and $\hat{\mathcal{W}}^{(s)}_{(\Delta, \pm 1)}$ to the two-point correlation function $\langle F_{\mu \nu}(x) F_{\alpha \beta}(x')  \rangle$. Since $F_{\mu \nu}$ is antisymmetric, we only have six independent components in four dimensions, namely $F_{w \bar{w}}, F_{wz}, F_{w \bar{z}}, F_{\bar{w} z}, F_{\bar{w} \bar{z}}, F_{z \bar{z}}$.  We will focus on the components $\langle F_{w \bar w}(x) F_{w \bar w} (x') \rangle$ and 
$\langle F_{z \bar z}(x) F_{z \bar z} (x') \rangle$ which are enough to determine the contribution of the conformal block to both  $g_1(\chi, \xi_2)$ and $H(\chi)$.
Then by applying \eqref{g2g3g5}, we can figure out the contributions to $g_2$, $g_3$, $g_4$ and $g_5$, and hence to the complete
$\langle F_{\mu \nu}(x) F_{\alpha \beta}(x')  \rangle$ two-point function.

Focusing then on these two components, we find
\be
\label{FwwFww}
\langle F_{w \bar w}(x) F_{w \bar w} (x') \rangle &=& - \frac{1}{4 |z|^2 |z'|^2} g_1 - \frac{|w-w'|^2}{8|z|^3 |z'|^3} \frac{\partial g_1}{\partial \chi}
\ , \\
\label{FzzFzz}
\langle F_{z \bar z}(x) F_{z \bar z} (x') \rangle &=&  \frac{1}{4 |z|^2 |z'|^2} g_1 + \frac{|w-w'|^2}{8|z|^3 |z'|^3} \frac{\partial g_1}{\partial \chi}  \\
&&
+ \frac{|w-w'|^2 - |z|^2 - |z'|^2}{16|z|^3 |z'|^3} H + \frac{|w-w'|^4 - (|z|^2 - |z'|^2)^2}{32|z|^4 |z'|^4} H' \ . \nonumber
\ee
Interestingly $H(\chi)$, because it lacks $\theta$ dependence, can only contribute to conformal blocks with $s=0$.  In fact, 
the $G_0$ mode of $g_1$ and $H(\chi)$ contribute in a linearly dependent way to $\langle F_{z \bar z}(x) F_{z \bar z} (x') \rangle$.
For the $s \neq 0$ modes, we find that 
$\left. \langle F_{z \bar z}(x) F_{z \bar z} (x') \rangle\right|_{s \neq 0} = - \left. \langle F_{w \bar w}(x) F_{w \bar w} (x') \rangle \right|_{s \neq 0}$, something we will see born
out by the conformal block contributions to these components of the correlation function.

Likewise, the component $\langle F_{w \bar w}(x) F_{z \bar z}(x')\rangle$ is enough to determine the contribution of the conformal block to the parity odd function $\tilde{g}_1(\chi)$. Combining the solution to the PDE's \eqref{g2g3g5} (with the replacement $g_i \rightarrow \tilde{g}_i$) and Bose symmetry \eqref{parityoddBose} we find that $\tilde{H}(\chi)$ is fixed,  
\begin{equation}
\tilde{H}(\chi) = 2 \partial_\chi  \left( (\chi^2 - 1) \tilde g_1(\chi)  \right) \ .
\end{equation}
The required component is
\be
\label{FwwFzz}
\langle F_{w \bar w}(x) F_{z \bar z} (x') \rangle &=& \frac{1}{4 |z|^2 |z'|^2} \tilde{g}_1 + \frac{|w-w'|^2}{8|z|^3 |z'|^3} \frac{\partial \tilde{g}_1}{\partial \chi}
\ . 
\ee

\subsection*{Vector Contributions}

The expectation, which will be born out, is that the conformal block contributions of $\hat \psi_{\pm}^{(\pm s)}$ 
and $\hat \phi_{\pm}^{(\pm s)}$ correspond to the modes
$G_{s}$ in the decomposition of $g_1$.

We start by looking at $\langle F_{w \bar w}(x) F_{w \bar w}(x') \rangle$. Applying both the operators ${\mathcal A}_{\mu\nu}$ and ${\mathcal B}_{\mu\nu}$ to the two point function $\langle \hat{\psi}_+^{(s)} \hat{\psi}_-^{(-s)}  \rangle$ and both ${\mathcal C}_{\mu\nu}$ and ${\mathcal D}_{\mu\nu}$ to $\langle \hat{\phi}_+^{(s)} \hat{\phi}_-^{(-s)}  \rangle$,
we can write the contribution of $\hat \psi_\pm^{(s)}$ and $\hat \phi_\pm^{(s)} $to the $\langle FF \rangle$ two point function as
\begin{equation} \label{Maxwellblockspin}
\begin{aligned}
\left. \langle F_{w \bar{w}}(\omega) F_{w \bar{w}}(\omega') \rangle\right|_s = \bigg[
&-\frac{1}{4}\left(c^{(s)}_{+-} + \bar{c}^{(s)}_{+-} \right)[(z \bar{z}')^s + (\bar{z} z')^s]  \\
&\sum_{m,n} \frac{(-|z|^2)^m (-|z'|^2)^n}{m! n! (s)_{m+1} (s)_{n+1}} \frac{(2+s)_{m+n} (s)_{m+n+2}}{(|w-w'|^2)^{2+s+m+n}} \bigg] \ .
\end{aligned}
\end{equation}
(Note this expression is the contribution from both $+s$ and $-s$ modes, assuming $s>0$.) 
With some care, this double sum evaluates to a hypergeometric function (see (\ref{diffopdiffop})):
\begin{equation}
\label{FwwFwwhyper}
\left. \langle F_{w \bar{w}}(\omega) F_{w \bar{w}}(\omega') \rangle\right|_s 
= -\frac{c_{\mathfrak{R}}^{(s)}\cos(s\theta)}{s(s+1)}  \frac{\partial_w \partial_{\bar w}}{|z| |z'|} \frac{{}_2F_1\left(\frac{1+s}{2}, \frac{2+s}{2}; 1+s; \frac{1}{\chi^2} \right)}{(2\chi)^{1+s}} \ ,
\end{equation}
where $c_{\mathfrak{R}}^{(s)} = \mathfrak{Re}[c^{(s)}_{+-}]$. The fact that the $s=0$ defect vectors do not contribute to the conformal block decomposition of $\langle FF \rangle$ is reflected here in the divergence
of the expression at $s=0$.

We need now to compare this expression (\ref{FwwFwwhyper}) with (\ref{FwwFww}).  To that end, note that a general function of $\chi$ expands out under the operation of $\partial_w \partial_{\bar w}$ as 
\begin{equation}
\partial_w \partial_{\bar{w}}F(\chi) = \frac{\partial F}{\partial \chi} \frac{1}{2|z||z'|} + \frac{\partial^2F}{\partial \chi^2} \frac{|w-w'|^2}{4|z|^2|z'|^2}.
\end{equation}
We deduce that
\begin{equation} \label{spinningg1block}
\begin{aligned}
g_1|_s &= \frac{2 c_{\mathfrak{R}}^{(s)} \cos(s\theta)}{s(s+1)} \partial_\chi  \left[\frac{{}_2F_1\left(\frac{1+s}{2}, \frac{2+s}{2}; 1+s; \frac{1}{\chi^2} \right)}{(2\chi)^{1+s}}  \right] \ , \\
&=  \frac{4 c_{\mathfrak{R}}^{(s)} \cos(s\theta)}{s(s+1)}
G_s(\chi) \ .
\end{aligned}
\end{equation}
A similar calculation looking at $\langle F_{z \bar z}(x) F_{z \bar z}(x') \rangle$ gives the same result for $g_1$, with no
added contribution from $H$, as expected. 
In the language of conformal blocks, the ${\mathbb Z}_m$ orbifold theory considered in section \ref{orbifold}, in the region
$r>1$,  would have $c_{\mathfrak{R}}^{(s)} = -8 m s(s+1)$ if $s \in m {\mathbb Z}$ and zero otherwise. 

To determine the parity odd sector and the $\tilde g_i$ functions, 
we start by looking at $\langle F_{w \bar w}(x) F_{z \bar z}(x') \rangle$. Following similar steps as for the parity even case, we can write the contribution from $\hat{\psi}_{\pm}^{(\pm s)}$ and $\hat{\phi}_{\pm}^{(\pm s)}$ to the two point function as
\begin{equation}
\langle F_{w \bar w}(x) F_{z \bar z }(x') \rangle|_s \sim \frac{c_{\mathfrak{I}}^{(s)} \sin(s\theta)}{s(s+1)} \frac{\partial_w \partial_{\bar w}}{|z| |z'|} \frac{{}_2F_1\left(\frac{1+s}{2}, \frac{2+s}{2}; 1+s; \frac{1}{\chi^2} \right)}{(2\chi)^{1+s}},
\end{equation}
where $c_{\mathfrak{I}}^{(s)} = \mathfrak{Im}[c^{(s)}_{+-}]$. Comparing with \eqref{FwwFzz} gives the conformal block contribution to the parity odd function $\tilde{g}_1$, 
\begin{equation}
\tilde{g}_1|_s \sim \frac{2 c_{\mathfrak{I}}^{(s)} \sin(s\theta)}{s(s+1)} G_{s}(\chi). 
\end{equation}
However, Bose symmetry \eqref{parityoddBose} requires that $\tilde{g}_3 = 0$ and hence combining with \eqref{g2g3g5} we find $c_{\mathfrak{I}}^{(s)} = 0$ for each $s > 0$.   Thus, the contribution from a defect vector to $\tilde{g}_1|_s$ vanishes completely. Only the $s=0$ scalars have the possibility to contribute to $\tilde{g}_1$ and the parity odd sector.

\subsection*{Scalar Contributions}

To compute the scalar conformal block contribution to $\langle FF \rangle$, we first need to compute the action of
the differential operator ${\mathcal D}^2$ on the defect scalar two-point functions.
Using the result (\ref{DolanOsborn}), we can reduce the double sum ${\mathcal D}^2$ to a single sum, which in turn can be expressed as a hypergeometric function.
In fact, we find the $s=0$ version of (\ref{FwwFwwhyper}) but without the leading divergent $1/s (s+1)$ coefficient:
\begin{equation}
\begin{aligned}
&\mathcal{D} \left[ \mathcal{D}' \left[ \frac{1}{|w-w'|^4} \right] \right] =  \frac{1}{2 |z||z'|}  \partial_w \partial_{\bar{w}} \frac{1}{\sqrt{-1+\chi^2}}
 \ .
 \end{aligned}
 \end{equation}

 As for the defect vector case, we can study the components $\langle F_{w \bar w} F_{w \bar w} \rangle$, $\langle F_{z \bar z} F_{z \bar z} \rangle$, and $\langle F_{w \bar w} F_{z \bar z} \rangle$ to find the conformal block contribution from the defect scalars to the $\langle FF \rangle$ two-point function. We deduce that the scalar $s=0$ contribution to $g_1$, $H$, and $\tilde g_1$ is 
 \begin{equation} \label{scalarblockbroken}
g_1|_{s=0} = - \chi \frac{c_{\parscalarnohat \parscalarnohat}}{4(\chi^2-1)^{\frac{3}{2}}}, \quad H(\chi) =   \frac{c_{\parscalarnohat \parscalarnohat} + c_{\perpscalarnohat  \perpscalarnohat}}{4(\chi^2-1)^{\frac{3}{2}}}, \quad 
\tilde{g}_1|_{s=0} =\chi \frac{c_{\parscalarnohat \perpscalarnohat}}{4(\chi^2-1)^{\frac{3}{2}}} \ .
\end{equation}
Note that in terms of the mode functions $g_1|_{s=0} = c_{\parscalarnohat \parscalarnohat} G_0(\chi)$ and
$H(\chi) = -(c_{\parscalarnohat \parscalarnohat} + c_{\perpscalarnohat  \perpscalarnohat}) G_1(\chi)$
(see footnote \ref{fn:Gchi}).
We see directly that breaking of the parity symmetry (non-zero $c_{\parscalarnohat \perpscalarnohat}$) is required to obtain a contribution to $\tilde{g}_1$.

\section{Bulk-Defect-Defect Correlation Functions}
\label{sec:FWW}
\begin{table}[]
	\centering
	\begin{tabular}{|clcccc|}
		\hline
		\hline
		\hline
		\multicolumn{6}{|c|}{\textbf{Bulk-Defect-Defect $U(1)$ Structures}}                                                                                                                     \\ \hline
		\hline
		\hline
		\multicolumn{2}{|c|}{At $x_1$} & \multicolumn{2}{c|}{At $\textbf{x}_2$}                                            & \multicolumn{2}{c|}{At $\textbf{x}_3$}                        \\ \hline
		\multicolumn{2}{|c|}{$\ell_k=s_k=0$}    & \multicolumn{1}{c|}{$\ell_2$}          & \multicolumn{1}{c|}{$s_2$}                   & \multicolumn{1}{c|}{$\ell_3$}          & $s_3$                    \\ \hline
		\multicolumn{2}{|c|}{$\hat{\mathcal{V}}^{(1)}_{\mu}, \hat{\mathcal{V}}^{(2)}_{\mu}, \hat{\mathcal{V}}^{(3)}_{\mu}, \hat{\mathcal{V}}^{(4)}_{\mu}$}         & \multicolumn{1}{c|}{$T_2^{\ell_2}$} & \multicolumn{1}{c|}{$\left(\frac{z}{|z|} \right)^{s_2}$} & \multicolumn{1}{c|}{$T_3^{\ell_3}$} & $\left(\frac{z}{|z|}\right)^{s_3}$ \\ \hline \hline \hline
	\end{tabular}
	\caption{A table containing all the independent tensor structures appearing in a Bulk-Defect-Defect three-point function. Here $\ell_k$ and $s_k$ are the parallel and transverse spin respectively of a defect operator inserted at $\textbf{x}_k$ and all the bulk structures have zero defect spin. The definition of each of these structures can be found in \eqref{bulkdefectdefectU1}. There is also trace removing Kronecker delta.}
	\label{bulkdefectdefectU1table}
\end{table}

In this section, by applying the free field constraints and the defect OPE to the bulk-defect-defect correlation function of $F_{\mu\nu}$ and two defect operators,
we place strong constraints on the defect operator spectrum of the theory.
A defect operator $\hat {\mathcal W}_i({\bf x})$ will have quantum number $s_i$, $\ell_i$, and $\Delta_i$.
As the defect operators live in two dimensions, it will be useful
to introduce the conformal weights $h'+ h = \Delta$ and $h'- h = \ell$. 

Our principal character is
\be
\label{bulkdefectdefectdef}
\left \langle  F_{\mu \nu}(x_1)  \hat {\mathcal W}_2({\bf x}_2) \hat {\mathcal W}_3({\bf x}_3)  \right \rangle \ .
\ee
This bulk-defect-defect three-point function picks out the operators in the defect OPE of $F_{\mu\nu}$ which have spin $s_1=-s = -s_2 -s_3$, by angular momentum conservation.  Because of the free field constraints, there are exactly two operators which contribute.  When $s<0$, they are $\hat \psi_+^{(-s)}$ and $\hat \phi_+^{(-s)}$.  When $s>0$,  they are correspondingly $\hat \psi_-^{(-s)}$ and $\hat \phi_-^{(-s)}$.  
Finally, when $s=0$, we have scalar contributions from $\parscalar$ and $\perpscalar$.  
We label by $h_1$ and $h'_1$ the conformal weights of the operators $\{ \hat \psi_\pm^{( s_1)}, \hat \phi_\pm^{( s_1)}, \parscalar, \perpscalar \}$ in the defect OPE of $F_{\mu\nu}$.   

The end result of this section is that for the bulk-defect-defect three point function to be well behaved, the weights $h_i$ 
and $h'_i$ must satisfy a ``double twist'' condition.  Without loss of generality, let us assume $h_3 \geq h_2$.\footnote{%
We make this assumption for the simplicity of stating the result. No such assumption is used in the analysis presented in this section.
}  Then 
we find
\be
\label{doubletwist}
h_3 = h_1 + h_2 + n \ , \; \; \; h'_3 = h'_1 + h'_2 + n'  \ ,
\ee
where $n$ and $n'$ are non-negative integers.  A further corollary is that 
for $\hat  {\mathcal W}_1 \in \{ \hat \psi_\pm^{( s_1)}, \hat \phi_\pm^{( s_1)}, \parscalar, \perpscalar \}$
the corresponding three point functions,
\be
\left \langle \hat  {\mathcal W}_1({\bf x}_1)  \hat {\mathcal W}_2({\bf x}_2) \hat {\mathcal W}_3({\bf x}_3)  \right \rangle \ ,
\ee
must vanish unless this double twist condition (\ref{doubletwist}) is satisfied (along with the constraint $s_1 + s_2 + s_3 = 0$).

\subsection{Sketch of the Approach}

The obvious approach to a proof would be to apply the differential operators ${\mathcal A}_{\mu\nu}$, ${\mathcal B}_{\mu\nu}$, ${\mathcal C}_{\mu\nu}$, ${\mathcal D}_{\mu\nu}$, $D^{(\tau)}_{\mu\nu}$ and $D^{(\pi)}_{\mu\nu}$ (partially defined in (\ref{ABdefs}), (\ref{CDdefs}), and (\ref{Ddef}))
to a defect three-point function
to generate the corresponding contributions to $\langle F_{\mu\nu} \hat {\mathcal W}_2 \hat {\mathcal W}_3 \rangle$.  We ran into technical obstacles
with this approach, however.  
A trivial objection is that the differential operators ${\mathcal A}_{\mu\nu}$ etc.\
were made explicit only for the $z \bar z$ and $w \bar w$ 
components of the three-point function.  In fact, we can easily generate the operators that give us the remaining
tensorial components of $\langle F_{\mu\nu} \hat {\mathcal W}_2 \hat {\mathcal W}_3 \rangle$, but have spared the reader
the details.  A more substantial obstacle 
is that we were not able to perform the sum in full generality.  In the scalar case,  the authors of 
\cite{Lauria:2020emq} get around this 
obstacle by evaluating the sum in a particular limit and recovering the general result using conformal invariance.
Unfortunately, because of the tensorial nature of our three-point function,
the constraints from conformal invariance are  less trivial and need to be understood in detail to use this approach.

As a first step, we find the most general form for $\langle F_{\mu\nu} \hat {\mathcal W}_2 \hat {\mathcal W}_3 \rangle$
consistent with conformal invariance.  Then instead of returning to the defect OPE, 
we apply the free field constraints directly to the full 
$\langle F_{\mu\nu} \hat {\mathcal W}_2 \hat {\mathcal W}_3 \rangle$ to constrain its form.  In effect, the equation of motion and Bianchi
identity convert the computation of the defect OPE sum to the solution of a set of differential equations.
The boundary conditions are then fixed by the defect OPE.  Thus
 in a final step, we return to the approach of \cite{Lauria:2020emq}.  
 We complete the calculation of $\langle F_{\mu\nu} \hat {\mathcal W}_2 \hat {\mathcal W}_3 \rangle$
by comparing the result from the free field constraints with the action of the differential operators 
${\mathcal A}_{\mu\nu}$, ${\mathcal B}_{\mu\nu}$, ${\mathcal C}_{\mu\nu}$ and ${\mathcal D}_{\mu\nu}$ on the defect three-point function, evaluated in a particular limit where we can perform the sum, but where now the system is constrained
enough that this limit uniquely determines the form of $\langle F_{\mu\nu} \hat {\mathcal W}_2 \hat {\mathcal W}_3 \rangle$.

\subsection{Conformal Invariance and Free Field Constraints}

This three-point function $\langle F_{\mu\nu} \hat {\mathcal W}_2 \hat {\mathcal W}_3 \rangle$ 
can depend on functions of a cross ratio $u$ \cite{Lauria:2017wav}, which we can write as
\be
u = \frac{({\bf x}_2 - x_1)^2 ({\bf x}_3-x_1)^2}{({\bf x}_2-{\bf x}_3)^2 |y_1|^2} = \frac{(w_{21} \bar w_{21} + z_1 \bar z_1) (w_{31} \bar w_{31} + z_1 \bar z_1)}{w_{23} \bar w_{23} z_1 \bar z_1}  \ .
\ee
Note the $u \to \infty$ limit is both the limit in which $x_2$ and $x_3$ become coincident and also separately the limit in which 
$x_1$ approaches the defect.  
The undetermined functions $f_A(u)$ are summed over a finite number of tensor structures.
As we discuss in more detail in Appendix \ref{app:bulkdefectdefect}, there are six tensor structures that contribute to the correlation function,
\be
\label{fullsolFWW}
 \left \langle  F_{\mu\nu}(x_1)  \hat {\mathcal W}_2({\bf x}_2)  \hat {\mathcal W}_3({\bf x}_3)  \right \rangle  = 
 {\mathcal N} \sum_{A=1}^6 f_A(u) S^A_{\mu\nu}(x_1, {\bf x}_2, {\bf x}_3) \ ,
\ee
where the prefactor 
\be
{\mathcal N} = \frac{\left( \frac{z_1}{|z_1|} \right)^s |z_1|^{ \Delta_2 + \Delta_3 -2} \, T_2^{\ell_2} T_3^{\ell_3}}{(w_{12} \bar w_{12} + z_1 \bar z_1)^{\Delta_2} 
(w_{13} \bar w_{13} + z_1 \bar z_1)^{\Delta_3} }
\ee
encodes the angular momentum $s$ and $\ell_i$ dependence of the defect operators 
$\hat {\mathcal W}_{(\Delta_i, \ell_i)}^{(s_i)}({\bf x}_i)$. 
 Note $s = s_2 + s_3$. 
The denominator in the prefactor along with the $|z_1|$
dependence in the numerator guarantee that the overall expression has the correct $\Delta_2 + \Delta_3 + 2$ scaling weight under dilatations.  
The weight zero tensor structures $S_{\mu\nu}^A$ (see \eqref{definingSmunuA}) are formed by the antisymmetric product of the basis vectors $\hat{\mathcal{V}}^{(i)}_{\mu}$ which are described in detail in Appendix \ref{app:bulkdefectdefect}.  The tangential angular momentum ``tensors'' are
\be \label{theTs}
T_2 = \frac{w_{12}}{4 |z_1|} - \frac{w_{12} \bar w_{12} + z_1 \bar z_1}{ 4|z_1| w_{23} \bar w_{23}} w_{32} \ , \; \; \;
T_3 = \frac{w_{13}}{ 4|z_1|} - \frac{w_{13} \bar w_{13} + z_1 \bar z_1}{ 4|z_1| w_{23} \bar w_{23}} w_{23}  \ .
\ee

To apply constraints on this three-point function, we assume that $F_{\mu\nu}$ obeys the Maxwell equations of motion and the Bianchi identity.
These eight equations are not all linearly independent, but allow us to solve for the $f_A$ up to several integration constants. The precise details of the solution depend sensitively on the choice of parameters $\Delta_i$, $\ell_i$, and $s$.
We treat the main cases $s\neq 0$ and $s=0$ below.  Several special cases are treated separately in appendix \ref{app:specialcases}.

\subsubsection*{Case: $s \neq 0$ and $\ell_2 \neq \ell_3$}

To begin, consider the case $\ell_2 \neq \ell_3$ and $s \neq 0$.  
Solving the four equations of motion $\partial_\mu F^{\mu\nu} = 0$ lets us solve for $f_3$, $f_5$ and $f_6$ in terms of the remaining $f_A$:
\be 
f_3 &=& -\frac{1}{2us} \left(8 (1 + \Delta_2 - (h_2+h_3) u) f_1 + (1 + \Delta_2 - 2 h_2 u) f_2 + u(u-1) (8 f_1' + f_2') \right) \ , 
\nonumber 
\\
f_5 &=& \frac{1}{2 u s} ( 8 (1 + \Delta_3 - (h_2 + h_3) u) f_1 + f_2  + (u (2h_2 - 1) - \Delta_2)  f_4  
\nonumber
 \\
\label{f356rel}
&& \hspace{1in}- u f_2  + u (u-1) (8 f_1' - f_4') )\ , \\
f_6 &=& \frac{4}{us} \left( (\Delta_3 - (h_2 + h_3 -1) f_2 + (\Delta_2 + u - (h_2 + h_3) u) f_4 + u(u-1) (f_2' + f_4') \right) \ . 
\nonumber
\ee
The Bianchi identity $\partial_\mu F_{\lambda \rho} \epsilon^{\lambda \rho \mu \nu} = 0$ then produces a relation 
\be
\label{f1rel}
f_1 &=& \frac{1}{8 (\ell_3 - \ell_2)} \bigl( (\ell_2 + 2 h_3(1-u) + u) f_2 - (\Delta_2 + (1-2h_2) u) f_4 \nonumber \\
&& \hspace{1in} + u (u-1) (f_2' - f_4') \bigr) \ .
\ee
Note in particular the denominators $s$ and $\ell_2 - \ell_3$ which will force us to consider the cases 
$s=0$ and $\ell_2 = \ell_3$ separately.  That the case $s=0$ needs to be considered separately is perhaps
not surprising because in this case scalars in the defect OPE of $F_{\mu\nu}$, not vectors, contribute 
to the three point function.

The four relations (\ref{f356rel}) and (\ref{f1rel}) can be used to simplify the original set of ODEs
to a pair of decoupled second order ODE's, one for $f_2$ and one for $f_4$.  Defining a new variable $v = \frac{1}{1-u}$, 
these two ODE's have hypergeometric solutions:\footnote{%
\label{fn:id}
The identity
$
 {}_2 F_1\left(a,b,c,v\right)  = (1- v)^{c-a-b} {}_2 F_1\left(c-a,c-b,c, v\right)  
$
is useful in solving these ODEs.
}
\be
\begin{split}
f_2 &=& \frac{(1-v)^{\Delta_2+1}}{v^{h_2 + h_3 - 1}} \biggl( c_1 v^{-\frac{s}{2}} {}_2 F_1\left(1+ h_2 - h_3 - \frac{s}{2}, 1+ h'_2 -h'_3 - \frac{s}{2}, 1-s, v\right)  +
 \\
&& +  c_2 v^{\frac{s}{2}} {}_2 F_1 \left( 1+ h_2 - h_3 +\frac{s}{2}, 1+ h'_2 -h'_3 + \frac{s}{2}, 1+s, v \right)  \biggr)  \ , \\
f_4 &=& \frac{(1-v)^{\Delta_3+1}}{v^{h_2 + h_3 - 1}} \biggl( c_3 v^{-\frac{s}{2}} {}_2 F_1\left(1+ h_3 - h_2 - \frac{s}{2}, 1+ h'_3 -h'_2 - \frac{s}{2}, 1-s, v\right)  +  
 \\
&& + c_4 v^{\frac{s}{2}} {}_2 F_1 \left( 1+ h_3 - h_2 + \frac{s}{2}, 1+  h'_3 -h'_2 +\frac{s}{2}, 1+s, v \right) \biggr)\ .
\end{split}
\ee
In order to have good behavior in the defect limit $v \to 0$, 
we expect to be able to set $c_1 = 0 = c_3$ when $s>0$.  Indeed, a more careful comparison
with the defect OPE justifies this choice a posteriori.  Allowing $c_1$ and $c_3$ to be nonzero would be equivalent
to allowing operators below the unitarity bound into the defect OPE of $F_{\mu\nu}$.  
Note when $s<0$, we could use the opposite pair of solutions and set $c_2 = 0 = c_4$ instead. In general, then, we expect the two solutions
 \be
 \label{f2f4sols}
 \begin{split}
 f_2 &=& c_2 \frac{(1-v)^{\Delta_2+1}}{ (-v)^{h_2 + h_3 -1- \frac{|s|}{2}} }{}_2 F_1 \left( 1+ h_2 - h_3 +\frac{|s|}{2}, 1+ h'_2 -h'_3 + \frac{|s|}{2}, 1+|s|, v \right)   \ , \\
f_4 &=& c_4 \frac{(1-v)^{\Delta_3+1} }{ (-v)^{h_2 + h_3-1 - \frac{|s|}{2}}} {}_2 F_1 \left( 1+ h_3 - h_2 + \frac{|s|}{2}, 1+  h'_3 -h'_2 +\frac{|s|}{2}, 1+|s|, v \right) \ .
\end{split}
\ee

We can give physical meaning to the integration constants $c_2$ and $c_4$ by analyzing the defect OPE
of $F_{\mu\nu}$.
Angular momentum conservation means only operators in the defect OPE with transverse spin $-s = -s_2 - s_3$ will contribute to the three point function.
Restricting to the case $s \neq 0$, 
only the operators with parallel spin $\ell=\pm 1$ contribute:
\be
 \lefteqn{ \left \langle  F_{\mu \nu}(x_1) \hat  {\mathcal W}_2({\bf x}_2) \hat {\mathcal W}_3({\bf x}_3)  \right \rangle
 = } \\
 &&
 \hspace{0.5in} \begin{cases}
\left \langle \left (  {\mathcal A}_{\mu\nu}^{(s)}\hat \psi_+^{(-s)}  + {\mathcal C}_{\mu\nu}^{(s)} \hat \phi_+^{(-s)}\right )\!({\bf x}_1)\, \hat {\mathcal W}_2({\bf x}_2) \hat {\mathcal W}_3({\bf x}_3)  \right \rangle  
 &   s<0 \\
 \left \langle\left (  {\mathcal B}_{\mu\nu}^{(s)}\hat \psi_-^{(-s)}  + {\mathcal D}_{\mu\nu}^{(s)} \hat \phi_-^{(-s)}\right )\!({\bf x}_1)  \, \hat {\mathcal W}_2({\bf x}_2) \hat {\mathcal W}_3({\bf x}_3) \right \rangle 
 & s> 0\nonumber 
 \end{cases}
\ . \nonumber
 \ee
We can isolate the contributions from the $\ell=1$ and $\ell=-1$ modes by looking at specific tensorial components of the three-point function. 
For example, from (\ref{vanishingABCD}), we see that the $\ell=1$ and $s<0$ modes are the only contribution to the $F_{wz}$ component, while the 
$\ell=-1$ and $s<0$ are the only contribution to the $F_{w \bar z}$ component.  In general, the integration constants $c_2$ and $c_4$ map to specific linear combinations 
of the three point function coefficients $c_{\psi 23}$ and $c_{\phi 23}$.  
Through evaluating the defect OPE in the large $w_2$ and large $w_3$ limits, it is possible to establish that
\be
\begin{split}
c_2 &=& \frac{(-1)^{-\ell_3} 2^{2(\ell_3 - \ell_2+1) }}{|s|} \left( c_{\psi 23} \left( h_2 -  h_3 + \frac{|s|}{2} \right) -c_{\phi 23} \left( h'_2 -  h'_3 + \frac{|s|}{2}\right)\right) \ , \\
c_4 &=& \frac{(-1)^{-\ell_2} 2^{2(\ell_3 - \ell_2+1) }}{|s|} \left(  c_{\psi 23} \left( h_2 -  h_3 -\frac{|s|}{2} \right) - c_{\phi 23} \left( h'_2 -  h'_3 - \frac{|s|}{2}\right)\right) \ .
\end{split}
\ee

Now there remains a subtle issue with this three point function, which is that it is not well behaved everywhere it should be.  The hypergeometric functions have
singular behavior at three special points, $u = 0$, 1 and $\infty$.  In our case, $u = 0$ is not achievable for physical locations of the insertions, and we thus
have no intuition of a condition to impose there.  On the other hand $u \to \infty$ corresponds both to the coincident limit $w_2 \to w_3$ and the defect limit
$|z_1| \to 0$.  Here, it makes sense that the behavior can be singular, but we have already selected boundary conditions to make sure
$f_2$ and $f_4$ are compatible with the defect OPE.  Finally, $u=1$ corresponds to a whole locus of points where $x_1$ lies on a semicircle perpendicular to the boundary with the line joining the boundary points $x_2$ and $x_3$ as the diameter.  
For this set of points, we do not generically expect singular behavior in the correlation function even though the
hypergeometric functions may well be singular there.  

Let us examine in more detail configurations with $u\approx 1$. More specifically, we set $w_2 = -w_3 = 1$, $w_1 = 0$, and $z_1 = 1 + \epsilon$ with $\epsilon \in {\mathbb R}$. 
Schematically, we find for $s>0$,
\be
\label{wzexp}
\lefteqn{
\langle F_{wz}(x_1) \hat {\mathcal W}_2({\bf x}_2) \hat {\mathcal W}_3({\bf x}_3) \rangle \sim (u-1)^{\frac{\ell_3-\ell_2-1}{2}}
 \frac{ c_{\phi 23} \Gamma(1+\ell_2 - \ell_3) \Gamma(|s|+1)}
{\Gamma\left(1+h_3-h_2 + \frac{|s|}{2} \right) \Gamma\left (h'_2- h'_3 + \frac{|s|}{2} \right) } 
}  \nonumber \\
&& \hspace{1.1in} +   (u-1)^{\frac{\ell_2-\ell_3-1}{2}}
 \frac{ c_{\phi 23}\Gamma(1+\ell_3 - \ell_2) \Gamma(|s|+1)}
{\Gamma\left(1+h_2-h_3 + \frac{|s|}{2} \right) \Gamma\left (h'_3- h'_2 + \frac{|s|}{2} \right) } \ , \\
\label{wbzexp}
\lefteqn{
\langle F_{w\bar z}(x_1) \hat {\mathcal W}_2({\bf x}_2) \hat {\mathcal W}_3({\bf x}_3) \rangle \sim
  (u-1)^{\frac{\ell_3-\ell_2-1}{2}}
 \frac{ c_{\psi 23}\Gamma(1+\ell_2 - \ell_3) \Gamma(|s|+1)}
{\Gamma\left(h_3-h_2 + \frac{|s|}{2} \right) \Gamma\left (1+h'_2- h'_3 + \frac{|s|}{2} \right) } 
}
\nonumber   \\
&& \hspace{1.1in} +  (u-1)^{\frac{\ell_2-\ell_3-1}{2}}
 \frac{  c_{\psi 23}\Gamma(1+\ell_3 - \ell_2) \Gamma(|s|+1)}
{\Gamma\left(h_2-h_3 + \frac{|s|}{2} \right) \Gamma\left (1+h'_3- h'_2 + \frac{|s|}{2} \right) }  \ .
\ee
(For $s<0$, the expansion is the same, but $F_{wz}$ and $F_{w \bar z}$ change places.)
To get this expansion, it is important to recognize that the factors $T_2$ and $T_3$ are also potentially singular in the $u \to 1$ limit,
as can be seen from the fact that
$T_i \bar T_i = \frac{1}{16}(u-1)$.  
The expression is schematic in nature.  The hypergeometric functions expand as double power series in $(u-1)$, and we have given only the 
leading term in each respective power series.

Clearly either $\ell_2 - \ell_3$ or $\ell_3 - \ell_2$ is negative, which will mean that generically one of the two terms in the $u=1$ expansions above will diverge because of its $u \approx 1$ behavior.
 In our particular
case $\ell_2 - \ell_3$ is an integer.  Given that $\ell_2 \neq \ell_3$ (the special case $\ell_2 = \ell_3$ will be treated below), the $-1/2$ in the exponent is not sufficient to render the other term singular.  There is a further consequence from $\ell_2 - \ell_3$ being integer: the two power series will overlap, and there will be logarithms in the expansion near $u=1$ that start at order $\frac{|\ell_2 - \ell_3|}{2}$.  The diverging
$\Gamma$ function in the supposedly finite term is an indication that there are logarithms present at this order in the actual power series
expansion in this case.  
Nevertheless, we can learn
something from a consideration of this schematic expression.

To remove the leading divergent term, the expressions (\ref{wzexp}) and (\ref{wbzexp}) suggest that one of the
 $\Gamma$-functions in the denominator be evaluated at a pole.  In fact, 
choosing the argument of any of these $\Gamma$-functions to be a negative integer means that the hypergeometric
functions (\ref{f2f4sols}) will have a polynomial type form, that the leading divergent series expansion will be completely absent, and that the logarithms
that start at order $\frac{|\ell_2 - \ell_3|}{2}$ will be removed, saving the situation and rendering
the $u=1$ limit of the correlation function finite.\footnote{%
 Recall that a hypergeometric function ${}_2 F_1 (-n, b, c, z)$ with a negative integer index $n = 0, 1, 2, \ldots$ has a polynomial form.
Further, a hypergeometric of the form
 ${}_2 F_1 (a, b, b-n, z)$ can be reduced to polynomial form through the identity 
 in footnote \ref{fn:id}. 
}
The only other option is to set the constants $c_{\phi 23}$ and $c_{\psi 23}$ to zero,
which sets $\langle F_{\mu\nu} \hat{\mathcal W}_2 \hat{\mathcal W}_3 \rangle$ to zero as well.

We argue that these four possible conditions on the $\Gamma$-functions in the denominator imply a ``double twist'' condition on defect operators.   
In particular, let $h_1$ and $h_1'$ be the conformal weights of 
one of the $\hat \psi^{(s_1)}_\pm$ or $\hat \phi^{(s_1)}_\pm$ operators in the defect OPE of $F_{\mu\nu}$.  Then $h_2 = h_1 + h_3 + n$ and $h'_2 = h'_1 + h'_3 + n'$ where $n$ and $n'$ are non-negative integers, or alternatively $h_3 = h_1 + h_2 + n$ and $h'_3 = h'_1 + h'_2 + n'$.

A more detailed argument follows. 
\begin{itemize}

 \item
For the $\ell=-1$ case, we have $h_1 = \frac{|s|}{2}+1$ and $h'_1 = \frac{|s|}{2}$.  
In the case $\ell_3 < \ell_2$ or equivalently $h_2 - h'_2 < h_3 - h'_3$,
then $h_3 - h_2 + \frac{|s|}{2} + 1 = -n$ or $h'_2 - h'_3 + \frac{|s|}{2}  = -n'$.
Equivalently $h_2 = h_1 + h_3 + n$ or $h'_3 = h'_1 + h'_2 + n'$.  
Now given a condition on $h'_3$, we want to see if there is any implication for $h_3$.  
Using that $h'_3 = h'_1 + h'_2 + n'$, we find that $h_3 > h'_1 +h_2 + n' = h_1 + h_2 + n' -1$.  
In other words $h_3 = h_1 + h_2 + n-1$ where in order for $\ell_2$ and $\ell_3$ to be integer, 
$n$ must be an integer such that $n > n'$.
Using instead that $h_2 = h_1 + h_3 + n $, we find that $h'_2 > h_1 + h'_3 + n = h'_1 + h'_3 + n+1$.  
In other words, $h'_2 = h'_1 + h'_3 + n'+1$ where $n'$ is an integer such that $n' > n$.  

For   $\ell_3 > \ell_2$, we have the same argument with $2 \leftrightarrow 3$.

\item
For the $\ell=1$ case, we have $h_1 = \frac{|s|}{2} $ and $h'_1 = \frac{|s|}{2}+1$.   
In the case $\ell_3 < \ell_2$ or equivalently $h_2 - h'_2 < h_3 - h'_3$,
then $h_3 - h_2 + \frac{|s|}{2}  = -n$ or $h'_2 - h'_3 + \frac{|s|}{2} +1 = -n'$ .
Equivalently $h_2 = h_1 + h_3 + n$ or $h'_3 = h'_1 + h'_2 + n'$.  
Using that $h'_3 = h'_1 + h'_2 + n'$, we find that $h_3 > h'_1 +h_2 + n' = h_1 + h_2 + n' +1$.  
In other words $h_3 = h_1 + h_2 + n+1$ where $n$ is an integer such that $n > n'$.
Using that $h_2 = h_1 + h_3 + n$, we find that $h'_2 > h_1 + h'_3 + n = h'_1 + h'_3 + n-1$.
In other words $h'_2 = h'_1 + h'_3 + n' - 1$ where $n' > n$.  

For   $\ell_3 > \ell_2$, we have the same argument with $2 \leftrightarrow 3$.

\end{itemize}

\subsubsection*{The scalar case: $s=0$}

When
$s=0$,  the system of constraints on the bulk-defect-defect correlation function
breaks apart into separate sets of equations for $f_1$, $f_2$, $f_4$ and for $f_3$, $f_5$, $f_6$.

For the $f_3$, $f_5$, $f_6$ system, it is convenient to introduce the combination $f_{35}(u) = f_3(u) - f_5(u)$.  One finds
\be
\label{s0f5eq}
f_5 &=& \frac{(-\ell_2 + 2 h_3(u-1) )f_{35}(u) +(1-u) u f_{35}'(u)}{2 (h_2 - h_3) (u-1)} \ , \\
f_6 &=& \frac{4 (\ell_3 - \ell_2) f_{35}(u)}{(h_3-h_2) (u-1)} \ .
\ee
along with a second order differential equation for $f_{35}(u)$, whose solution is 
\be
f_{35}(u)&=& (u-1)^{-\ell_2} ( c_{35} u^{\Delta_2}  {}_2 F_1(h_2 - h_3, 1+h_2-h_3, 1+h_2-h_3 + h'_2-h'_3, u)  \nonumber \\
&& + c_{35}' u^{\Delta_3}  {}_2 F_1 (h'_3 - h'_2, 1+h'_3 - h'_2, 1-h_2 + h_3 -h'_2+h'_3, u) 
) \ .
\ee
To eliminate logarithms from the solution in the defect limit $u \to \infty$, the constants $c_{35}$ and $c_{35}'$ can 
be adjusted to give
\be
f_{35}  =  \tilde c_{35} (u-1)^{-\ell_2}  u^{h_3 + h'_2 - 1}  {}_2 F_1 \left( 1 - h_2' + h_3', 1+h_2-h_3, 2, \frac{1}{u} \right) \ .
\ee

For the $f_1$, $f_2$, $f_4$ system, we find instead
\be
\label{s0f1eq}
f_1 &=& \frac{(\ell_3 - \ell_2 + h_2 (u-1) -h_3 u) f_2 + (h_2 - h_3) u f_4}{8(\Delta_2 - \Delta_3)} \\
\label{s0f4eq}
f_4 &=& \frac{\Delta_2 - \Delta_3}{(h_2 - h_3)(h'_2 - h'_3) u} \biggl( \left(  -  \Delta_3+ 
\left(2 h_3 - 1 + \frac{(h_2-h_3)^2}{\Delta_2 - \Delta_3} \right) u \right) f_2 \nonumber \\
&& -(u-1)u f_2' \biggr) \ .
\ee
Now $f_2$ satisfies a second order ODE whose general solution is a hypergeometric function
\be
f_2 &=& (u-1)^{-1 - \ell_3} \biggl( c_2 u^{\Delta_3} {}_2 F_1 (h_3 - h_2, h_3 - h_2, h_3 -h_2 + h'_3 - h'_2, u) \nonumber
\\
&& + c_2' u^{1+ \Delta_2} {}_2 F_1 ( 1 + h'_2 - h'_3, 1+h'_2 - h'_3, 2 + h_2 - h_3 + h'_2 - h'_3, u) \biggr) \ .
\ee
Again to eliminate logarithms from the solution in the defect limit $u \to \infty$, one finds
\be
f_2 &=& \tilde c_2 (u-1)^{-1-\ell_3} u^{h_2 + h_3'} {}_2 F_1 \left( h_3 - h_2, 1+h_2' - h_3', 1, \frac{1}{u} \right) \ .  
\ee

To separate out the $\hat \tau$ and $\hat \pi$ contributions to the three-point function, we can examine particular tensor components
to which one or the other scalar do not contribute.  In particular, comparing with (\ref{Drels}), 
it is clear that $\hat \tau$ contributes only to the $z \bar z$ component and $\hat \pi$ contributes only to the $w \bar w$ component.  Comparing with the large $w_2$ and $w_3$ limits of the defect OPE, it is possible to 
establish that $\tilde c_2 \sim c_{\pi 2 3}$ while $\tilde c_{35} \sim (h_2 - h_3) c_{\tau 23}$.  
We find for $s=0$ and $\ell_2 \neq \ell_3$ 
that near $u = 1$, 
\be
\label{sequal}
\frac{1}{c_{\tau 23}} \langle F_{z \bar z}(x_1) \hat{\mathcal W}_2({\bf x}_2) \hat{\mathcal W}_3({\bf x}_3) \rangle &\sim& \frac{1}{c_{\pi 23}} \langle F_{w \bar w}(x_1) \hat{\mathcal W}_2({\bf x}_2) \hat{\mathcal W}_3({\bf x}_3) \rangle \\
&\sim &
(u-1)^{\frac{\ell_3 - \ell_2-1}{2}}
  \frac{\Gamma(\ell_2 - \ell_3+1) }{\Gamma \left( h_3 - h_2  +1 \right) \Gamma \left( h'_2 - h'_3 + 1\right) }\nonumber  \\
  &&
\pm  (u-1)^{\frac{\ell_2-\ell_3-1}{2}} 
   \frac{  \Gamma(\ell_3 - \ell_2+1)}{\Gamma \left( h_2 - h_3 +1 \right) \Gamma \left( h'_3 - h'_2 +1 \right) } \ ,  \nonumber
\ee
which leads to the same double twist condition on the defect spectrum.

In more detail:

\begin{itemize}

\item 
For the scalar, we have $h_1 = h'_1 = 1$.  
In the case $\ell_3 < \ell_2$ or equivalently $h_2 - h'_2 < h_3 - h'_3$,  then $h_3 - h_2 +1 =-n$ or 
$h'_2 - h'_3  +1 = -n'$.  
Equivalently, we can write this constraint
as $h_2 = h_1 + h_3 +n$ or $h'_3 = h'_1  +h'_2 + n'$.
Now given a condition on $h'_3$, we want to see if there is any implication for $h_3$.  
Note that since $h_3 - h_2 > h'_3 - h'_2$, we find that $h_3 > h'_1 + h_2 + n' = h_1 + h_2 + n'$.  
If we further assume an integer spin condition, i.e.\
that $h_i - h'_i$ is integer, then we find that $h_3 = h_1 + h_2 + n$ for $n$ some non-negative integer $n > n'$.  
We can play the same game with $h'_2$ and the condition $h_2 = h_1 + h_3 + n$.  

In the case $\ell_3 > \ell_2$, 
 we apply  the same argument as above with $2 \leftrightarrow 3$.

\end{itemize}

There unfortunately remain a number of special cases which need separate analysis to finish the proof.
These are choices of parameters where the set of equations we solved above become singular.
When $\ell_2 = \ell_3$, the equation (\ref{f1rel}) for $f_1$ becomes singular and needs special treatment.
In looking at the $\ell_2 = \ell_3$ case in greater detail, we will find further that the subcase
$\Delta_2 - \Delta_3 = \pm s$
needs special treatment as well.
When $s=0$ and $\Delta_2 = \Delta_3$, the equation (\ref{s0f1eq}) for $f_1$ becomes singular.  
When either $h_2 = h_3$ or $h'_2 = h'_3$, the equations (\ref{s0f5eq}) and (\ref{s0f4eq}) for $f_4$ and $f_5$ become
singular.  Indeed the cases $h_2 = h_3$, $h'_2 = h'_3$ and $\Delta_2 - \Delta_3 = \pm s$ are
 particularly troublesome because the free field constraints seem to lead
to solutions which are well behaved at $u=1$, which naively would require a relaxation of the double twist constraint
to include some negative integers.  A careful analysis of the defect OPE constraint rules out
these possibilities, however.  We describe these cases in appendix \ref{app:specialcases}.
At the end of the day, we find the double twist conditions that $h_3 = h_1 + h_2 + n$ and $h'_3 = h'_1 + h'_2 + n'$ where
$n$ and $n'$ are non-negative integers and we have assumed $h_3 > h_2$.  

To sum up, the analysis presented in this section (and in appendix \ref{app:specialcases}) shows that the finiteness of $\langle F_{\mu \nu}(x_1) \hat{\mathcal{W}}_2(\textbf{x}_2) \hat{\mathcal{W}}_3(\textbf{x}_3) \rangle$ as $u \rightarrow1$ implies either \{$\hat{\mathcal{W}}_2$ is a double-twist combination of $\hat{\mathcal{O}}$ and $\hat{\mathcal{W}}_3$\} OR \{$\hat{\mathcal{W}}_3$ is a double-twist combination of $\hat{\mathcal{O}}$ and $\hat{\mathcal{W}}_2$\}, where $\hat{\mathcal{O}}$ is any operator appearing in the defect OPE of $F_{\mu \nu}$.
From this result, we have the following claim.
\paragraph{Claim:} The Operator Product Expansion between any two operators $\hat{\mathcal{O}}_1, \hat{\mathcal{O}}_2$ appearing in the defect OPE of $F_{\mu \nu}$ has to be of the form, 
\begin{equation}
\hat{\mathcal{O}}_1 \times \hat{\mathcal{O}}_2 \sim \mathbf{1} + \{\hat{\mathcal{W}}_k \, | \,  h_k = h_1 + h_2 +n, \,  h_k' = h_1'+h_2'+n' \, | \, n, n' \in \mathbb{N}  \}.
\end{equation}
\paragraph{Proof:} Here we will write $\hat{\mathcal{W}}|_{\hat{\mathcal{O}}_1 \hat{\mathcal{O}}_2}$ to mean that $\hat{\mathcal{W}}$ is a double-twist combination of $\hat{\mathcal{O}}_1$ and $\hat{\mathcal{O}}_2$. Then, let's consider $\langle F_{\mu \nu}(x_1) \hat{\mathcal{O}}_2(\textbf{x}_2) \hat{\mathcal{W}}(\textbf{x}_3) \rangle$ and $\langle F_{\mu \nu}(x_2) \hat{\mathcal{O}}_1(\textbf{x}_1) \hat{\mathcal{W}}(\textbf{x}_3) \rangle$. From the arguments in this section we obtain the conditions \{$\hat{\mathcal{W}}|_{\hat{\mathcal{O}}_1 \hat{\mathcal{O}}_2}$ OR $\hat{\mathcal{O}}_2 |_{\hat{\mathcal{O}}_1 \hat{\mathcal{W}}}$ \} AND \{$\hat{\mathcal{W}}|_{\hat{\mathcal{O}}_1 \hat{\mathcal{O}}_2}$OR $\hat{\mathcal{O}}_1|_{\hat{\mathcal{O}}_2 \hat{\mathcal{W}}}$\}. Out of the four possible conditions we see that only $\hat{\mathcal{W}}|_{\hat{\mathcal{O}}_1 \hat{\mathcal{O}}_2}$ is allowed and thus conclude the OPE $\hat{\mathcal{O}}_1 \times \hat{\mathcal{O}}_2$ must only contain double twist operators. 
The other conditions lead to the requirement of double twist on more than one operator and hence leads to a contradiction. For example, consider $\hat{\mathcal{W}}|_{\hat{\mathcal{O}}_1 \hat{\mathcal{O}}_2}$ AND $\hat{\mathcal{O}}_1|_{\hat{\mathcal{O}}_2 \hat{\mathcal{W}}}$. This means $h_3 = h_1 + h_2 + m$ and $h_1 = h_2 + h_3 + n$, which implies $h_2 + m + n = 0$ (same for $h'$). Since reflection positivity demands $h_2 \geq 0$ and as $m,n \in \mathbb{N}$, this condition cannot be satisfied (expect for the identity).

\section{Triviality of the Defect}
\label{sec:trivial}

The result and proof in this section are a variant of the result and proof of ref.\ \cite{Lauria:2020emq},
where the authors consider a set of scalar operators in a conformal field theory
in more than one dimension.  They show that if the scalar operators have a generalized free field spectrum,
then all the $n$-point functions are of generalized free fields as well.  Here, we specialize to a two dimensional
conformal field theory but consider operators of any spin.  In particular, we assume that in the operator product expansion,
\be
{\mathcal W}_1{\mathcal W}_2   = \delta_{h_1, h_2} \delta_{h'_1, h'_2} {\bf 1} + 
\sum_k {\mathcal W}_k \ ,
\ee
we find only operators with conformal weights $h_k = h_1 + h_2 + n$ and $h'_k = h'_1 + h'_2 + n'$ with $n$, $n' \in {\mathbb N}$.  
In the context of the earlier sections, we are working here purely on the two dimensional defect, ignoring the rest of the space-time.
Thus these operators ${\mathcal W}_i$ would be defect operators $\hat {\mathcal W}_i$ in earlier sections although we drop the `hat' here.
\\

\noindent
The main ingredient of the proof is this restriction on the operator product expansion of two such generalized free fields:
\be
 {\mathcal W}_1(\textbf{x})  {\mathcal W}_2 (0,0) = \frac{\delta_{h_1, h_2} \delta_{h'_1, h'_2}}{w^{h_1 + h_2} \bar w^{h'_1 + h'_2} } {\bf 1}
+ \sum_k \frac{\lambda_{12k} }{w^{h_1 + h_2 - h_k} \bar w^{h'_1 + h'_2 - h'_k}} {\mathcal W}_k \ .
\ee
Due of the restriction of $h_k - h_1 - h_2$ and $h'_k - h'_1 - h'_2$ to non-negative integer values, the three-point function reduces to, 
\begin{equation}
\langle \mathcal{W}_1({\bf x}_1) \mathcal{W}_2 ({\bf x}_2) \mathcal{W}_k ({\bf x}_3) \rangle = \frac{\lambda_{12k} w_{12}^{n} \bar w_{12}^{n'} }{w_{13}^{h_1 + h_k -h_2}   \bar w_{13}^{h'_1 + h'_k - h'_2}   w_{23}^{h_2 + h_k - h_1}  \bar w_{23}^{h'_2 + h'_k - h'_1}   },
\end{equation}
and by Bose symmetry we find, 
\begin{equation}
\lambda_{12k} = (-1)^{n+n'} \lambda_{21k},
\end{equation}
without any branch point issues since $n$ and $n'$ are in $\mathbb N$. The identity term on the operator product expansion vanishes if $\langle \mathcal{W}_1 \mathcal{W}_2 \rangle = 0$ while it picks up a discontinuity when $\langle \mathcal{W}_1 \mathcal{W}_2 \rangle \neq 0$, associated with the branch point at $w=\bar w =0$ for non-integer exponents $h_1 + h_2$ and $h'_1 + h'_2$. Combining with the double-twist condition, we find that the commutator trivializes, 
\begin{equation}
\begin{aligned} \label{commrel}
[ {\mathcal W}_1(\textbf{x}) , {\mathcal W}_2(0,0) ] = \begin{cases} 
 \operatorname{Disc} \left[ \frac{\delta_{h_1, h_2} \delta_{h'_1, h'_2}}{w^{h_1 + h_2} \bar w^{h'_1 + h'_2} } \right] {\bf 1} & \langle \mathcal{W}_1 \mathcal{W}_2 \rangle \neq 0, \\
0 & \langle \mathcal{W}_1 \mathcal{W}_2 \rangle = 0.
\end{cases}
\end{aligned}
\end{equation}
\begin{figure}
\begin{center}
\includegraphics[width=4in]{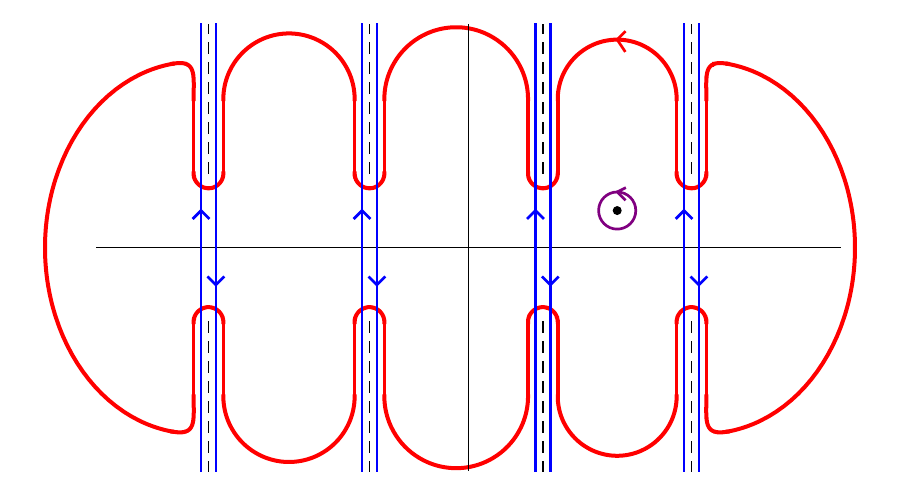}
\end{center}
\caption{The deformation of the contour used to produce the dispersion relation for a five-point function. 
We start with the small purple circle about the point $\tau$, then puff up
the contour into the red caterpillar shape, avoiding the branch cuts that start at $\tau_k \pm i y$, 
and finally trade the caterpillar for the series of vertical blue lines, one on each side
of the branch cuts, assuming, because of cluster decomposition, that we can discard the portions of the contour at infinity.}
\end{figure}

The proof of triviality proceeds inductively.  We know that $\langle {\mathcal W}_i(z_i, \bar z_i) \rangle$ vanishes while conformal invariance fixes
$\langle {\mathcal W}_i (z_i, \bar z_i) {\mathcal W}_j (z_j, \bar z_j) \rangle$ up to normalization.  If we assume Wick's Theorem holds for $(n-2)$-point
functions, it suffices then to show it holds for $n$-point functions.  To that end, 
consider the $n$-point function
\be
G_n(\tau) = \langle {\mathcal W}_1 (w_1, \bar w_1)  {\mathcal W}_2(w_2, \bar w_2)  \cdots  {\mathcal W}_n (w_n, \bar w_n) \rangle \ .
\ee
It is convenient to begin by choosing special locations for the points $(w_i, \bar w_i)$.
We place all but the first operator on the real line, $\bar w_k = w_k = \tau_k \in {\mathbb R}$ for $2 \leq k \leq n$, and we order them such that $\tau_{k-1} > \tau_k$.  The first operator we place at $(w_1 = \tau + i y, \bar w_1 = \tau - i y)$ where $\tau \in {\mathbb C}$.  The correlation function is analytic
except along cuts starting at $\tau = \tau_k \pm i y$ and running off to $\pm i \infty$. 
We then write the correlation function as a dispersion relation:
\be
G_n(\tau) &=& \oint \frac{d \tau'}{2 \pi i} \frac{G_n(\tau')}{\tau'-\tau} 
 =\int_{-\infty}^\infty \frac{dt'}{2 \pi} \times \\ 
 && \times \biggl(
\frac{1}{\tau - \tau_2 -i t'} \langle [ {\mathcal W}_1(\tau_2 +i t'+ i y, \tau_2 + i t' - iy), {\mathcal W}_2(\tau_2, \tau_2)] 
{\mathcal W}_3(\tau_3, \tau_3) \cdots {\mathcal W}_n(\tau_n, \tau_n) \rangle \nonumber \\
&& + 
\frac{1}{\tau - \tau_3 -i t'} \langle {\mathcal W}_2(\tau_2, \tau_2) [ {\mathcal W}_1(\tau_3 +i t' + i y, \tau_3 + i t' - iy), {\mathcal W}_3(\tau_3, \tau_3)] 
 \cdots {\mathcal W}_n(\tau_n, \tau_n) \rangle \nonumber \\
 &&+ \cdots \nonumber  \\
&& + \frac{1}{\tau - \tau_n -i t'} \langle {\mathcal W}_2(\tau_2, \tau_2) {\mathcal W}_3(\tau_3, \tau_3) \cdots [ {\mathcal W}_1(\tau_n +i t' + i y, \tau_n + i t' - iy), {\mathcal W}_n(\tau_n, \tau_n)]  \rangle \biggr) \nonumber \
\ee
Expanding the original small contour around $\tau$ to instead run up and down the cuts that start at $\tau_j \pm i y$, we have dropped pieces of the contour
at large $|\tau|$.  Dropping these pieces relies on an assumption of cluster decomposition.
If we insert the result for the commutator (\ref{commrel}) into this expression, some standard contour integral manipulations reduce the integrals to
a sum over $n-1$ $(n-2)$-point functions:
\be
\label{Wickresult}
G_n(\tau) &=& \frac{\delta_{h_1, h_2} \delta_{h'_1, h'_2}}{z_{12}^{h_1 + h_2} \bar z_{12}^{h'_1 + h'_2}} 
\langle {\mathcal W}_3(\tau_3, \tau_3) {\mathcal W}_4(\tau_4, \tau_4) \cdots {\mathcal W}_n(\tau_n, \tau_n) \rangle \\
&& + \frac{\delta_{h_1, h_3} \delta_{h'_1, h'_3}}{z_{13}^{h_1 + h_3} \bar z_{13}^{h'_1 + h'_3}} 
\langle {\mathcal W}_2(\tau_2, \tau_2) {\mathcal W}_4(\tau_4, \tau_4) \cdots {\mathcal W}_n(\tau_n, \tau_n) \rangle + \ldots \nonumber \\
&& +\frac{\delta_{h_1, h_n} \delta_{h'_1, h'_n}}{z_{1n}^{h_1 + h_n} \bar z_{1n}^{h'_1 + h'_n}} 
\langle {\mathcal W}_2(\tau_2, \tau_2) {\mathcal W}_3(\tau_3, \tau_3) \cdots {\mathcal W}_{n-1}(\tau_{n-1}, \tau_{n-1}) \rangle \ . \nonumber
\ee

The argument so far is for a particular choice of insertion points $z_i$ and $\bar z_i$.  To loosen this constraint, one can generalize the above proof to include not just primary operators ${\mathcal W}_i(z, \bar z)$ but their descendants as well, including potentially arbitrary numbers of derivatives acting on the primaries.
Thinking of ${\mathcal W}_i(z, \bar z)$ as a Taylor series expansion around ${\mathcal W}_i(\tau_i, \tau_i)$, this extension of the proof to descendant operators
allows us to consider general locations for the operator insertions.  

Starting with the vanishing of $\langle {\mathcal W}_i(w, \bar w) \rangle$ and a choice of normalization for the two-point functions
$\langle {\mathcal W}_i (w_i, \bar w_i) {\mathcal W}_j (w_j, \bar w_j) \rangle$, we can build up arbitrary $n$-point functions inductively, using the relation (\ref{Wickresult}).  We find any $(2n+1)$-point function vanishes
while any $2n$-point function follows from Wick's Theorem.

Let us pause for a minute to consider what this result means for our Maxwell theory.  Let $\hat {\mathcal O}_i$ be defect operators that are present in the defect OPE of $F_{\mu\nu}$ and $\hat {\mathcal W}$ be any other defect operator in the theory.  We can conclude that
$\langle\hat  {\mathcal O}_1(w_1, \bar w_1) \cdots \hat {\mathcal O}_n (w_n, \bar w_n) \rangle$  follows from Wick's Theorem since
our double twist result implies that in any nonzero three-point function $\langle \hat {\mathcal O}_1 \hat {\mathcal O}_2 \hat {\mathcal W} \rangle$, 
the operator $\hat {\mathcal W}$ must satisfy the double twist condition.  
However, the moment we insert an operator 
$\hat {\mathcal W}$ that is not in the defect OPE of $F_{\mu\nu}$, we can no longer conclude that 
$\langle \hat {\mathcal O}_1(w_1, \bar w_1) \cdots \hat {\mathcal O}_n (w_n, \bar w_n) \hat {\mathcal W}(w, \bar w) \rangle$ follows from Wick's Theorem.  We must be able to assume the double-twist condition for the OPE of any two operators in this correlation function, and we cannot necessarily conclude it for the pair $\hat {\mathcal O}_i$ and  $\hat {\mathcal W}$.  
This exception is exploited by free scalar fields and allows for example the correlator $\langle \phi(x_1) \phi(x_2) {:} \phi^2(x_3) {:} \rangle$ to be nonzero.

We learn that the three point function
$\langle \hat {\mathcal O}_1 \hat {\mathcal O}_2 \hat {\mathcal O}_3 \rangle$ must vanish, and hence, summing over descendants,
so must 
\be
\langle F_{\mu \nu}(x_1)  \hat  {\mathcal O}_2(w_2, \bar w_2) \hat {\mathcal O}_3(w_3, \bar w_3) \rangle \ .  
\ee
 By this argument, of the bulk-defect-defect three point functions considered in section \ref{sec:FWW}, only 
  $\langle F_{\mu \nu}(x_1)   \hat {\mathcal W}_2(w_2, \bar w_2) \hat {\mathcal W}_3(w_3, \bar w_3) \rangle $ and
  $\langle F_{\mu \nu}(x_1)   \hat {\mathcal O}_2(w_2, \bar w_2) \hat {\mathcal W}_3(w_3, \bar w_3) \rangle $ will be nonzero,
  where $\hat {\mathcal O}_i$ is in the defect OPE of $F_{\mu\nu}$ but the $\hat {\mathcal W}_j$ are not.

\section{Discussion}

A motivation behind this work was to understand what happens when charged matter on a surface defect is coupled to a free Maxwell field in the bulk.
The conclusion here is that such a theory cannot preserve $SO(3,1) \times SO(2)$ defect conformal invariance (Euclidean).  We saw that the coupling between
any defect charge current and the bulk Maxwell field must vanish, 
$\langle F_{\mu\nu}(x) \hat {\mathcal W}_{(1, \pm 1)}^{(0)}({\bf x}') \rangle = 0$.  Indeed, to be consistent with defect conformal invariance, $F_{\mu\nu}$ can only couple 
on the defect to a dimension two scalar or a vector of dimension $ \Delta = 1 + |s|$ for $|s| > 0$.  Moreover, these defect operators in turn can only couple to
other defect operators whose spectrum obeys a `double twist' condition and hence behave like generalized free fields.

This result is  in line with the heuristic argument given in the introduction, that having coupled the bulk photons to defect charged matter, the effective propagator for the photon is a logarithm and thus automatically introduces a scale dependence to the physics.
There remains however an interesting physical problem: What are the correlation functions and transport properties of
 charged matter on a surface defect coupled to a bulk photon?
If the coupling is weak, the 
theory should have some remnant of conformal invariance, and the structures uncovered in this discussion may help in gaining control over the 
renormalization group flow away from the decoupled limit.  It would be interesting to see if the scale dependence of the
logarithmic solutions associated with the $G_0$ and $G_1$
conformal blocks, for example, might be related to the scale dependence of the effective photon propagator.

One curiosity about the Maxwell field $F_{\mu\nu}$ compared with free  scalars $\phi$ and fermions $\psi$
 is that the unitarity condition appears to place more stringent constraints on the bulk-defect function
 $\langle F_{\mu\nu}(x) \hat {\mathcal W}_{(\Delta, \ell)}^{(s)}({\bf x}') \rangle$ 
 than on $\langle \phi (x) \hat {\mathcal W}_{(\Delta, \ell)}^{(s)}({\bf x}') \rangle$
 or $\langle \psi(x) \hat {\mathcal W}_{(\Delta, \ell)}^{(s)}({\bf x}') \rangle$.
Applying the equations of motion to $F_{\mu\nu}$, we found the Maxwell field can only couple to a boundary scalar of $\Delta = 2$ or a 
vector of dimension $ \Delta = 1 \pm |s|$.  The unitarity bound for such a vector is $\Delta \geq 1$, giving just one choice of $\Delta$ for a given $s$.
 
In comparison, for a free bulk scalar in this 2d/4d setting, $\langle \phi(x) \hat {\mathcal W}_{(\Delta, \ell)}^{(s)}({\bf x}') \rangle$ is non-zero for $\ell=0$ and 
$\Delta = 1 \pm s$ \cite{Lauria:2020emq}.  Here however the unitarity bound is $ \Delta \geq 0$, which means that for transverse spin in the range $-1 < s < 1$, 
there is a possibility of having two defect fields with the same transverse spin but different dimension, provided we allow for non-integer $s$.  
Similarly, for the bulk free fermion, one anticipates \cite{Bianchi:2021snj} 
that it can couple to boundary $\ell = \pm \frac{1}{2}$ 
fermions with dimension $ \Delta = \frac{3}{2} \pm s$ while the unitarity
bound would imply $ \Delta \geq \frac{1}{2}$, again allowing for two defect operators with the same transverse spin but different dimension in the 
range $-1 < s < 1$.  Ref.\ \cite{Lauria:2020emq} argued that having the possibility of these two different dimension operators was key to being able 
to construct a nontrivial defect theory.  In the defect-defect-bulk three-point function, these extra conformal blocks allow for some extra freedom in avoiding
the spurious singularity at $u=1$.  Recall it was this singularity that forced us to restrict the spectrum to that of generalized free fields.
For the Maxwell field in this 2d/4d setting, it would seem to be impossible to have these extra solutions, even for noninteger $s$.

\section*{Acknowledgments}
We would like to thank  Edo Lauria, Andy Stergiou, and Balt van Rees for communication.
  C.~H. was supported in part by a Wolfson Fellowship from the Royal Society.
  This work was supported 
  by the U.K.\ Science \& Technology Facilities Council Grant ST/P000258/1.


\appendix

\section{Conformal Blocks}

We start with a series expansion result that is useful for deriving the differential operators that generate the conformal blocks:
\be
\label{usefulresult}
\frac{1}{(|w|^2 + |z|^2)^\alpha} &=& \frac{1}{|w|^{2\alpha}}\sum_{j=0}^\infty \frac{(\alpha)_j}{j! }  \left( - \left| \frac{z}{w} \right|^2 \right)^{j} \ .
\ee
We can use this result to rewrite the rational expression $w (|w|^2 + |z|^2)^{-s-2}$ that shows up in the $z \bar z$ and $w \bar w$ components of the
$\langle F_{\mu\nu}(x) \hat \psi_{\pm}^{(s)} \rangle$ bulk-defect two point functions:
\be
\label{diffops}
\frac{w}{( |w|^2 + |z|^2 )^{s+2} } &=& -\sum_{j=0}^\infty  \frac{1}{j! (s)_{j+1}} \partial_w^{j+1} \partial_{\bar w}^{j} \frac{(-|z|^2)^j }{w^{s} \bar w^{s+2}} \ .
\ee
From this expression, we can construct the differential operators ${\mathcal A}^{(s)}_{z \bar z}$, ${\mathcal A}^{(s)}_{w \bar w}$, ${\mathcal B}^{(s)}_{z \bar z}$ and
${\mathcal B}^{(s)}_{w \bar w}$ in (\ref{ABdefs}) in the text.  

The next step is to consider the action of two of these differential operators
\be
{\mathcal L} \equiv -(-|z|^2)^j \sum_{j=0}^\infty  \frac{1}{j! (s)_{j+1}} \partial_w^{j+1} \partial_{\bar w}^{j}
\ee
on a defect two point function $\langle \hat \psi_+^{(s)} (w) \hat \psi_-^{(-s)}(0) \rangle = w^{-s} \bar w^{-s-2}$:
\be
 {\mathcal L} {\mathcal L}' \frac{(w-w')^2}{|w-w'|^{2s+4}} &=&\sum_{m,n} \frac{(-|z|^2)^m (-|z'|^2)^n (2+s)_{m+n} (s)_{m+n+2}}{m! n! (s)_{m+1} (s)_{n+1} |w-w'|^{2(2+s+m+n)}} \ . 
\ee
To convert the double sum into a single sum, we make use of (B.8) of \cite{Dolan:2000ut}:
\be
\label{DolanOsborn}
\sum_{m,n = 0} \frac{1}{m! n!} \frac{(\lambda)_{m+n} (\kappa)_{m+n}}{(\kappa)_m (\kappa)_n}
 \frac{A^m B^n}{z^{2(\lambda+m+n)}} = \frac{1}{C^\lambda} \sum_{m=0} 
\frac{(\lambda)_{2m}}{m! (\kappa)_m} \left( \frac{AB}{C^2} \right)^m \ ,
\ee
where $z^2 = A+B+C$.
The final result is 
\be
\label{diffopdiffop}
 {\mathcal L} {\mathcal L}' \frac{(w-w')^2}{|w-w'|^{2s+4}} 
&=& \frac{(2 |z z'|)^{-(s+1)}}{s(1+s)} \partial_w \partial_{\bar w} \left[ \chi^{-(s+1)}  {}_2 F_1 \left( \frac{1+s}{2}, \frac{2+s}{2}, 1+s, \frac{1}{\chi^2} \right)  \right]
\ee
where
\[
\chi = \frac{|w-w'|^2 + z^2 + z'^2}{2 |z| |z'|} \ .
\]
A similar calculation holds for the $\ell = -1$ vectors and the defect scalars.

\section{$U(1)$ Tensor Structures}
\label{sec:U1structs}
Here we present in more detail the independent structures required for bulk-defect two-point functions and bulk-defect-defect three-point functions. 
\subsection{Bulk-Defect Structures}
Consider a bulk point $x = (w, \bar w, sz, \bar{z})$ and a defect point $\textbf{x}'=(w', \bar{w}')$. Then the $U(1)$ structures presented in table \ref{bulk-to-defect} are given by, 
\begin{equation} \label{bulkdefectU1}
\begin{aligned}
\hat{\Xi}^{(1)}_{\mu} &= \frac{|z|}{|\delta w|^2+|z|^2} \left(\delta \bar{w}, \delta w, \frac{\bar{z} (-|\delta w|^2 +|z|^2)}{2|z|^2}, \frac{z(-|\delta w|^2 +|z|^2)}{2|z|^2}  \right), \\
\hat{\mathcal{I}}_{\mu w} &= -\frac{1}{2}\frac{1}{|\delta w|^2 + |z|^2} \left(\delta \bar{w}^2, -|z|^2, \delta \bar{w} \bar{z}, \delta \bar{w} z   \right), \quad \hat{\mathcal{I}}_{\mu \bar z} = \frac{1}{4} \left(0,0,1, -\frac{z}{\bar z} \right)\\
\hat{\mathcal{I}}_{\mu \bar{w}} &= - \frac{1}{2} \frac{1}{|\delta w|^2 + |z|^2} \left(-|z|^2, \delta w^2, \delta w \bar{z}, \delta w z  \right),
\end{aligned}
\end{equation}
where $\delta w = w-w'$ etc. Writing $\frac{\bar z}{|z|}\hat{\mathcal{I}}_{\mu \bar z} = \hat{\mathcal{V}}_{\mu}^{(4)}$, all possible contractions yield
\begin{equation} \label{bulkdefectcontractionsU1}
\begin{aligned}
\hat{\Xi}^{(1)}_{\mu}\hat{\Xi}^{(1)\mu} = 1, \quad \hat{\mathcal{V}}^{(4)}_{\mu}\hat{\mathcal{V}}^{(4)\mu} = -\frac{1}{4}, \quad \hat{\mathcal{I}}_{\mu w}\hat{\mathcal{I}}_{\bar{w}}^{\mu} = \frac{1}{2}, \\
 \hat{\mathcal{I}}_{\mu w} \hat{\mathcal{I}}_{\nu \bar{w}} = \frac{1}{4}g_{\mu \nu}- \frac{1}{4}\hat{\Xi}^{(1)}_{\mu}\hat{\Xi}^{(1)}_{\nu}  + \hat{\mathcal{V}}^{(4)}_{\mu}\hat{\mathcal{V}}^{(4)}_{\nu} + \hat{\mathcal{I}}_{[\mu| w}\hat{\mathcal{I}}_{|\nu] \bar{w}},
\end{aligned}
\end{equation}
where the rest of the contractions yield zero and we had to introduce a new defect spin zero, bulk rank 2 tensor which is fully anti-symmetric, namely $\hat{\mathcal{I}}_{[\mu| w}\hat{\mathcal{I}}_{|\nu] \bar{w}}$ to ensure the set of tensor structures is closed under contractions and $U(1)$ product.
\subsection{Bulk-Defect-Defect Structures}\label{app:bulkdefectdefect}
Consider a bulk point $x_1 = (\textbf{x}_1, z, \bar{z})$ and two defect points $\textbf{x}_2, \textbf{x}_3$. Then the $U(1)$ structures presented in table \ref{bulkdefectdefectU1table} are given by,
\begin{equation} \label{bulkdefectdefectU1}
\begin{aligned}
&\hat{\mathcal{V}}^{(1)}_{\mu} = \frac{|z|}{|w_{12}|^2 + |z|^2} \left( \bar{w}_{12},  w_{12}, \frac{\bar{z} (-|w_{12}|^2+ |z|^2) }{2|z|^2}, \frac{z(-|w_{12}|^2+ |z|^2) }{2|z|^2} \right), 
\\
&\hat{\mathcal{V}}^{(2)}_{\mu} = \frac{|z|}{|w_{13}|^2 + |z|^2} \left( \bar{w}_{13},  w_{13}, \frac{\bar{z} (-|w_{13}|^2+ |z|^2) }{2|z|^2}, \frac{z(-|w_{13}|^2+ |z|^2) }{2|z|^2} \right), \\
&\hat{\mathcal{V}}^{(3)}_{\mu} = \frac{w_{12} \bar{w}_{13} + |z|^2}{8|z| w_{23}(|w_{13}|^2+|z|^2)} \left(-|z|^2, w_{13}^2, w_{13} \bar{z}, w_{13} z \right), \\ 
&\hat{\mathcal{V}}^{(4)}_{\mu} = \left(0,0,-\frac{\bar{z}}{4|z|}, \frac{z}{4|z|} \right),
\end{aligned}
\end{equation}
where we have already defined $T_2, T_3$ in \eqref{theTs}. All the possible contractions and $U(1)$ products are given by,
\begin{equation} \label{bulkdefectdefectcontractionsU1}
\begin{aligned}
\hat{\mathcal{V}}_{\mu}^{(m)} \hat{\mathcal{V}}^{(n)\mu} = \begin{pmatrix}
1 &1-\frac{2}{u} &\frac{1}{4} \left(1 - \frac{1}{u} \right) &0 \\
1-\frac{2}{u} &1 &0 &0 \\
\frac{1}{4} \left(1 - \frac{1}{u} \right) &0 &0 &0 \\
0 &0 &0 &-\frac{1}{4} 
\end{pmatrix}, \quad 
T_r \bar{T}_r = \frac{u-1}{16},
\end{aligned}
\end{equation}
where $r\in \{2,3\}$. Since we are in $d=4$ dimensions, the set of bulk vectors also forms a four dimensional vector space. As we have found four linearly independent bulk vectors, we conclude that it is a maximal basis set. Likewise, a correlation function involving a single $F_{\mu \nu}$ is constructed by antisymmetrising the product of these bulk vectors,
\begin{equation} \label{definingSmunuA}
S_{\mu \nu}^A = \sum_{(i,j) \in \mathcal{A}} \left(\hat{\mathcal{V}}^{(i)}_{\mu} \hat{\mathcal{V}}^{(j)}_{\nu} - \hat{\mathcal{V}}^{(i)}_{\nu} \hat{\mathcal{V}}^{(j)}_{\mu} \right),
\end{equation}
where $\mathcal{A} = \{(1,2), (1,3), (1,4), (2,3), (2,4), (3,4)\}$ and $A \in \{1, ... ,6 \}$ labels the elements in the order they appear in $\mathcal{A}$. This gives a total of six independent structures and again we have a maximal basis set as the dimension of $4 \times 4$ antisymmetric matrices is six. To explain how to derive these higher-point function tensor structures using the methods in ref.\ \cite{Herzog:2020bqw}, we have to take a detour and set up a more generalised approach, which will be 
the subject of appendix \ref{sec:bulkn}.
\section{Bulk $n$-Point Function}
\label{sec:bulkn}
Given $n$ bulk points $x_r=(\textbf{x}_r, y_r)$, we can construct a maximum (for large enough dimension) of $n(n-1)$ independent cross-ratios, 
\begin{equation}
\xi_1^{(x_r,x_s)} = \frac{x_{rs}^2}{4|y_r||y_s|}, \qquad \xi_2^{(x_r,x_s)} = \frac{y_r \cdot y_s}{|y_r| |y_s|},
\end{equation}
where $x_{rs} = x_r - x_s$. Then using the method outlined in ref.\ \cite{Herzog:2020bqw}, we find the following set of independent tensor structures, 
\begin{equation} \label{npointbulk}
\begin{aligned}
\Xi^{(1)(\partial x_r, x_s)}_{\mu} &= \frac{2|y_r|}{x_{rs}^2}(x_{rs})_{\mu} - (n_r)_{\mu}, \quad \Xi^{(2)(\partial x_r,x_s)}_{\mu} = \frac{(n_s)_{\mu}}{\xi_2^{(x_r,x_s)}} - (n_r)_{\mu}, \\ 
I_{\mu \nu}^{(\partial x_r, \partial x_s)} &= \delta_{\mu \nu} - \frac{2 (x_{rs})_{\mu}(x_{rs})_{\alpha}}{x_{rs}^2},\quad \mathcal{J}_{\mu \nu}^{'(\partial x_r,\partial x_s)} = \begin{cases} 
\delta_{ij}-\frac{(y_r)_j (y_s)_i}{y_r \cdot y_s} & \mu=i, \nu=j \\
0 & {\rm otherwise}
\end{cases} , \\
\mathcal{J}_{\mu \nu}^{(\partial^2 x_r)} &= \begin{cases} 
\delta_{ij}- (n_r)_i(n_r)_j & \mu=i, \nu=j \\
0 & {\rm otherwise}
\end{cases} ,
\end{aligned}
\end{equation}
where we use the notation $\partial^m x_r$ to indicate the object transforms like a $SO(d)$ rank-m tensor at $x_r$ and the order of indices match the order of points appearing for non-symmetric objects. Although the algebra has been omitted, it can be shown that the set of bulk $n$-point tensor structures is closed under contractions (similar to the two-point case \cite{Herzog:2020bqw}). We also introduce structures which have a finite defect limit and will be used in obtaining bulk-defect-defect tensor structures, 
\begin{equation}
\begin{aligned}
\Pi_{\mu}^{(x_r,x_s, \partial x_t)} &= \xi_1^{(x_r,x_t)} \left(\Xi^{(1)(x_s, \partial x_t)}_{\mu} - \Xi^{(1)(x_r, \partial x_t)}_{\mu} \right), \\ 
\mathcal{X}_{\mu}^{(x_r, \partial x_s)} &= \xi_2^{(x_r,x_s)} \left(\Xi^{(1)(x_r, \partial x_s)}_{\mu} - \Xi^{(2)(x_r, \partial x_s)}_{\mu} \right), \\
\mathcal{I}_{\mu \alpha}^{(\partial x_r, \partial x_s)} &= I_{\mu \alpha}^{(\partial x_r, \partial x_s)} - \Xi^{(1)(\partial x_r, x_s)}_{\mu} \mathcal{X}^{(x_r, \partial x_s)}_{\alpha}.
\end{aligned}
\end{equation}
We should emphasise that $\Pi_{\mu}^{(x_r,x_s,\partial x_t)}$, which is finite in the double limit
$y_s, y_t \rightarrow 0$, was not presented in ref.\ \cite{Herzog:2020bqw} since it is constructed using three points.
%

For a bulk-defect-defect three-point function, we need to take the boundary limit of several of the structure above. 
In this case we find the following set of generically independent tensor structures in general dimension, 
\begin{equation} \label{bulkdefectdefectstructures}
\begin{aligned}
&\hat{\Xi}^{(1)(\partial x_1, \textbf{x}_2)}_{\mu}, \quad \hat{\Xi}^{(1)(\partial x_1, \textbf{x}_3)}_{\mu}, \quad \hat{\mathcal{X}}_i^{(x_1, \partial \textbf{x}_2)}, \quad \hat{\mathcal{X}}_i^{(x_1, \partial \textbf{x}_3)}, \quad \hat{\Pi}_a^{(x_1,\partial \textbf{x}_2, \textbf{x}_3)}, \quad  \hat{\Pi}_a^{(x_1, \textbf{x}_2, \partial \textbf{x}_3)}, \\
&\hat{\mathcal{I}}_{\mu a}^{(\partial x_1, \partial \textbf{x}_2)}, \quad \hat{\mathcal{I}}_{\mu i}^{(\partial x_1, \partial \textbf{x}_2)}, \quad \hat{\mathcal{I}}_{\mu a}^{(\partial x_1, \partial \textbf{x}_3)},\quad \hat{\mathcal{I}}_{\mu i}^{(\partial x_1, \partial \textbf{x}_3)}, \quad \hat{I}_{ab}^{(\partial \textbf{x}_2, \partial \textbf{x}_3)}, \quad   \mathcal{J}_{\mu \nu}^{(\partial^2 x_1)} \ .
\end{aligned}
\end{equation}
We also can use Kronecker delta functions to remove traces.
The `hat' notation indicates the defect limit of the bulk tensor structure. The number of total structures and of each individual type matches the embedding space result in ref.\ \cite{Guha:2018snh}. Next we will explain how to simplify the notation by taking advantage of the fact that $SO(p) \times SO(q) = U(1) \times U(1)$ in the special case where
$p=q=2$.
\subsection*{$U(1)$ Defect Primary}
Let $\phi$ be a (continuous) conformal map on a $d$-dimensional Euclidean space with a $p$-dimensional flat defect. Then using $\phi$, we can obtain a position dependent $SO(d)$ element $(\mathcal{R}^{\phi}_d)^{\mu}_{\nu}(x) = \Omega_{\phi}(x) (\partial_{\nu}\phi^{\mu})(x)$ in the bulk. Likewise, we can construct position dependent $SO(p)$ and $SO(q)$ elements $(\mathcal{R}^{\phi}_d)^a_b(\textbf{x},0)$ and $(\mathcal{R}^{\phi}_d)^i_j(\textbf{x},0)$ respectively, on the defect. 
These elements enable us to define bulk primary fields as irreducible representations of $SO(d)$, and also to define defect primaries with parallel and transverse spin. 
\\

\noindent
In the special case when $d=4$ and $p=q=2$, the bulk primaries are real irreps of $SO(4)$ while the defect primaries are real irreps of $SO(2)$. However, since $SO(2) \sim U(1)$ we can instead consider the defect primaries to be complex irreps of $U(1)$, all of which are one dimensional and parametrised by integers.\footnote{%
 There can  be situations where the periodicity constraints should be loosened and rational or even real weights should be used, for example if there is a magnetic
 flux line threading the defect.
}
We write the parallel spin as $\ell$ and the transverse spin as $s$. Since these are complex irreps, it is best to work in complex coordinates. So we change into the coordinate $x=(w, \bar{w}, z, \bar{z})$ with the transition map $x(x_1,x_2,y_1,y_2) = (x_1+ix_2, x_1 - ix_2, y_1+iy_2, y_1-iy_2)$. Now the defect lies at $z=\bar{z}=0$. 
\subsection*{$SO(2)$ to $U(1)$ Tensor Structures} \label{SO2toU1}
A straightforward way to find the $U(1)$ tensor structures is by starting with the $SO(2)$ structures and changing into complex coordinates. The parallel or transverse $SO(2)$ index will transform into a diagonal $U(1)$ index.  
Without loss of generality, let's consider a defect vector $\hat{\mathcal{W}}_{\mu}$ in Cartesian coordinates with $\mu \in \{a,i \}$ (either parallel $a \in \{1,2\}$ or transverse $i \in \{3,4 \}$). After changing into complex coordinates we claim and use the convention that $\hat{\mathcal{W}}_{w}$ is a spin $\ell = -1$ parallel $U(1)$ vector, $\hat{\mathcal{W}}_{\bar{w}}$ is a spin $\ell = +1$ parallel $U(1)$ vector, $\hat{\mathcal{W}}_{z}$ is a spin $s = -1$ transverse $U(1)$ vector and $\hat{\mathcal{W}}_{\bar{z}}$ is a spin $s= +1$ transverse $U(1)$ vector, at the point which the transformed index is associated with. With this procedure, we can find a complete set of $U(1)$ tensor structures starting with the $SO(2)$ structures. 
\subsection{Bulk-Defect-Defect}
Let us now consider the bulk-defect-defect three-point function. Starting with \eqref{bulkdefectdefectstructures} and using the method just explained, we find the following list of $U(1)$ structures which, because of the reduced dimensionality, might no longer all be independent:
\begin{equation}
\begin{aligned}
&\hat{\Xi}^{(1)(\partial x_1, \textbf{x}_r)}_{\mu}, \quad \hat{\mathcal{X}}_z^{(x_1, \partial \textbf{x}_r)} \rightarrow -\frac{1}{2}\frac{\bar{z}}{|z|}, \quad \hat{\Pi}_w^{(x_1,\partial \textbf{x}_2, \textbf{x}_3)} \rightarrow \bar{T}_2,\quad  \hat{\Pi}_w^{(x_1, \textbf{x}_2, \partial \textbf{x}_3)} \rightarrow \bar{T}_3,  \\
&\hat{\mathcal{I}}_{\mu w}^{(\partial x_1, \partial \textbf{x}_r)}, \quad \hat{\mathcal{I}}_{\mu \bar{w}}^{(\partial x_1, \partial \textbf{x}_r)} , \quad \hat{\mathcal{I}}_{\mu z}^{(\partial x_1, \partial \textbf{x}_r)}, \quad \hat{\mathcal{I}}_{\mu \bar{z}}^{(\partial x_1, \partial \textbf{x}_r)}, \quad \hat{I}_{ww}^{(\partial \textbf{x}_2, \partial \textbf{x}_3)} \rightarrow \frac{8\bar{T}_2 \bar{T}_3}{u-1},
\end{aligned}
\end{equation}
where $r\in \{2,3\}$ and we also have $ \mathcal{J}_{\mu \nu}^{(\partial^2 x_1)}$. Note that the bar component is given simply by the complex conjugate of the component without bar.  
Note $\hat{I}_{w \bar w} = 0$.
\\

\noindent
So far the structures with bulk indices all have different defect spin at different points. As a remedy, we rescale the structures using either $T_2$, $T_3$, $\frac{z}{|z|}$ or $\frac{\bar{z}}{|z|}$ to make them defect spin zero. (Note $\mathcal{J}$ is already defect spin zero.)  
Rescaling yields the following set of defect spin zero bulk vectors, 
\begin{equation}
\begin{aligned}
\hat{\Xi}^{(1)(\partial x_1, \textbf{w}_r)}_{\mu}, \quad \hat{\mathcal{I}}_{\mu w}^{(\partial x_1, \partial \textbf{w}_r)} T_r, \quad \hat{\mathcal{I}}_{\mu \bar{w}}^{(\partial x_1, \partial \textbf{w}_r)} \bar{T}_r, \quad \frac{z}{|z|} \hat{\mathcal{I}}_{\mu z}^{(\partial x_1,  \partial \textbf{w}_r)} , \quad \frac{\bar{z}}{|z|} \hat{\mathcal{I}}_{\mu \bar{z}}^{(\partial x_1,  \partial \textbf{w}_r)},
\end{aligned}
\end{equation}
where $r \in \{2,3 \}$ and hence we have ten different bulk vectors. However, as expected in four dimensions $d=4$, not all of these are linearly independent. As mentioned earlier \eqref{bulkdefectdefectU1}, we can choose a basis set with four independent vectors, 
\begin{equation}
\begin{aligned}
\hat{\mathcal{V}}^{(1)}_{\mu} &:= \hat{\Xi}^{(1)(\partial x_1, \textbf{w}_2)}_{\mu}, \qquad 
\hat{\mathcal{V}}^{(3)}_{\mu} := \hat{\mathcal{I}}^{(\partial x_1, \partial \textbf{w}_3)}_{\mu \bar{w}}\bar{T}_3, \\ 
\hat{\mathcal{V}}^{(2)}_{\mu} &:= \hat{\Xi}^{(1)(\partial x_1, \textbf{w}_3)}_{\mu}, \qquad 
\hat{\mathcal{V}}^{(4)}_{\mu} := \frac{z}{|z|} \hat{\mathcal{I}}^{(\partial x_1, \partial \textbf{w}_3)}_{\mu z}, 
\end{aligned}
\end{equation}
using which any other bulk vector can be written as a linear combination. In fact, even the rank-2 tensor $\mathcal{J}$ is not independent, 
\begin{equation}
\mathcal{J}^{(\partial^2 x_1)}_{\mu \nu} = -4 \hat{\mathcal{V}}^{(4)}_{\mu} \hat{\mathcal{V}}^{(4)}_{\nu}.
\end{equation}
Since the set of independent tensor structures are closed under contraction and $U(1)$ product \eqref{bulkdefectdefectcontractionsU1}, we cannot form any new structures and hence have a complete set of basis structures required for constructing any bulk-defect-defect three-point correlation function. 
\subsection{Bulk-Defect}
Here we can start with the bulk-to-defect tensor structures in ref.\ \cite{Herzog:2020bqw}. Then following a similar approach to the bulk-defect-defect case, we find the following $U(1)$ structures:
\begin{equation}
\begin{aligned}
\hat{\Xi}^{(1)}_{\mu}, \quad \hat{\mathcal{X}}'_{z} \rightarrow -\frac{1}{2}\frac{\bar{z}}{|z|}, \quad \hat{\mathcal{I}}_{\mu w}, \quad \hat{\mathcal{I}}_{\mu \bar{w}}, \quad \hat{\mathcal{I}}_{\mu z}, \quad \hat{\mathcal{I}}_{\mu \bar{z}}, \quad \mathcal{J}_{\mu \nu}.
\end{aligned}
\end{equation}
For this case, we do not have a parallel spin object such as $T_r$ -- which was used to rescale and remove spin $\ell$ -- and hence $\hat{\mathcal{I}}_{\mu w}$ and $\hat{\mathcal{I}}_{\mu \bar{w}}$ are bi-vector type structures with parallel spin $\ell = -1$ and $ \ell =1$ respectively. We can still rescale the transverse spin however, and find 
\begin{equation}
\frac{\bar{z}}{|z|} \hat{\mathcal{I}}_{\mu \bar{z}} = -\frac{z}{|z|} \hat{\mathcal{I}}_{\mu z}, \qquad  \mathcal{J}_{\mu \nu} = - \frac{4z^2}{|z|^2} \hat{\mathcal{I}}_{\mu z} \hat{\mathcal{I}}_{\nu z},
\end{equation}
leaving us with just two linearly independent bulk vectors $\hat{\Xi}^{(1)}_{\mu}$ and $\frac{ \bar z}{|z|} \hat{\mathcal{I}}_{\mu \bar z}$. One subtlely here is that the product $\hat{\mathcal{I}}_{\mu w} \hat{\mathcal{I}}_{\nu \bar{w}}$ is a bulk rank-2 tensor with zero defect spin. It turns out that $\hat{\mathcal{I}}_{[\mu| w} \hat{\mathcal{I}}_{|\nu] \bar{w}}$ is linearly independent and cannot be written as the antisymmetrised product of the bulk vectors, and hence we have to add it to the set of tensor structures to form a set closed under $U(1)$ product and complete the basis set. A similar story happened for the bulk structures for two-point functions in ref.\  \cite{Herzog:2020bqw}, where the contraction of a bi-vector with itself yielded the independent rank-2 tensor $\mathcal{J}_{\mu \nu}$. 
In any event, since the set of independent tensor structures are closed under contraction and $U(1)$ product \eqref{bulkdefectcontractionsU1}, we cannot form any new structures and have a complete basis set required for constructing bulk-defect two-point correlation functions.

\section{Special Cases of the Bulk-Defect-Defect Correlation Function}
\label{app:specialcases}

\subsection{Case: $s \neq 0$: $\ell_2 = \ell_3$}

 The special case $\ell_2 = \ell_3$ and $s \neq 0$ requires some extra care.  
 The relations for $f_3$, $f_5$, and $f_6$ remain the same as in the $s\neq 0$, $\ell_2 \neq \ell_3$  case.  The relation for $f_1$ becomes a constraint on $f_2$ and $f_4$ leading to the solution
\be
\label{f4problem}
f_4 &=& \frac{\Delta_2 - \Delta_3}{u((\Delta_2 - \Delta_3)^2 - s^2)} \biggl(  - 4  (u-1) u f_2' + \\
&&
+ \left( -4 \Delta_3 + (-4 + \Delta_2 + 3 \Delta_3 - 4 \ell) u - \frac{u s^2}{\Delta_2 - \Delta_3} \right) f_2 \biggr) \ ,
\nonumber
\ee
and
\be
f_2 &=& c_2 \frac{(v-1)^{\Delta_3}}{ v^{\frac{\Delta_2 + \Delta_3}{2} -1- \ell - \frac{|s|}{2} }}
{}_2 F_1 \left( \frac{\Delta_3- \Delta_2 + |s|}{2},  \frac{\Delta_3- \Delta_2 + |s|}{2}, 1+|s|, v \right) \ .
\ee
Finally $f_1$ can be solved as a second order ODE with a source term
\be
\label{f1eq}
f_1'' + p(v) f_1' + q(v) f_1 = -\frac{f_2'}{8 v(1-v)} - \frac{(\Delta_2 + \ell (v-1) + v) f_2}{8 (v-1)^2 v^2} \ ,
\ee
where the homogeneous ($f_2 = 0$) solutions  are
\be
f_1 &=& \frac{(1-v)^{1 + \Delta_2}}{ v^{\frac{\Delta_2 + \Delta_3}{2} -\ell} }
\biggl( c_1 v^{-\frac{s}{2}} {}_2 F_1 \left( \frac{\Delta_2 - \Delta_3 - s}{2}, \frac{2 + \Delta_2 - \Delta_3-s}{2}, 1-s, v \right) \nonumber \\
&&
+ c_1' v^{\frac{s}{2}} {}_2 F_1 \left( \frac{\Delta_2 - \Delta_3 + s}{2}, \frac{2 + \Delta_2 - \Delta_3+s}{2}, 1+s, v \right) 
\biggr) \ .
\ee

Without solving for $f_1$, we note that there is already an issue associated with $f_2$, 
that it has logarithms near $u=1$:
\be
f_2 \sim
c \frac{(u-1)^{-1-\ell}  \log(u-1)}{\Gamma\left( 1 + \frac{\Delta_2 -\Delta_3 + |s|}{2} \right) \Gamma \left( 1+ \frac{\Delta_3 - \Delta_2 + |s|}{2} \right)} \ .
\ee
Observe that $\frac{\Delta_2 - \Delta_3}{2} = h_2 - h_3  = h'_2 - h'_3$ when $\ell_2 = \ell_3$.
The condition for the vanishing of these logarithms is then the same double twist condition
that we found in the $\ell_2 \neq \ell_3$ case above.  

Instead of solving (\ref{f1eq}) directly, the solution of interest can be isolated by studying the constraints from the defect OPE.  One finds
\be
f_1 &=& \frac{1}{16} \left( u - 1+ \frac{( 2 \alpha-1)u |s|}{\Delta_2 - \Delta_3} \right) f_2 + \frac{u}{16} \left( 1 - \frac{(2 \alpha-1)|s|}{\Delta_2 - \Delta_3} \right) f_4 \ ,
\ee
where
\be
\alpha = \frac{c_{\phi 23}}{c_{\phi 23} + c_{\psi 23}}
\ee
for $s<0$ and we swap $c_{\phi 23} $ and $c_{\psi 23}$ for $s>0$.
The integration constant $c_2 \sim (c_{\phi 23} + c_{\psi 23}) (\Delta_2 - \Delta_3 + |s|)$ on the other hand.  
We find that for $s<0$ and $u \approx 1$,
\be
\frac{\langle F_{wz} \hat {\mathcal W}_2 \hat {\mathcal W}_3 \rangle }{c_{\phi 23}}\sim 
\frac{ \langle F_{w \bar z} \hat {\mathcal W}_2 \hat {\mathcal W}_3 \rangle }{c_{\psi 23}}
\sim \frac{s! (u-1)^{-1/2} }{\Gamma \left(\frac{\Delta_2 - \Delta_3+|s| }{2} +1 \right)
\Gamma \left(\frac{\Delta_3- \Delta_2+|s|}{2} +1 \right)} \  .
\ee
Thus the logarithm disappears at leading order (but will show its head at subleading order), and finiteness does indeed require
the double twist condition.  (There is a similar expression for $s>0$, but with the $wz$ and $w \bar z$ components swapped.)

\subsection{Case: $\ell_2 = \ell_3$, $\Delta_2 - \Delta_3 = s > 0$}
This case corresponds to $h_2 = h_3 + \frac{s}{2}$ and $h'_2 = h'_3 + \frac{s}{2}$.  While the free field constraints
suggest that there is a solution with the right boundary conditions, we will find this solution is incompatible with the defect OPE.
The solution from the free field constraints is
\begin{subequations}
\be
f_1 &=& c_1 \frac{u^{\Delta_3 + 1}}{(u-1)^{\ell+1}} + c'_1 \frac{u^{\Delta_2+1}}{(u-1)^{\ell+1}} -\frac{c_2}{8} \frac{u^{\Delta_3}}{(u-1)^{\ell+1}} \ , \\
f_2 &=& c_2 \frac{u^{\Delta_3}}{(u-1)^{\ell+1}} \ , \\
f_3 &=& -2 c'_1 \frac{u^{\Delta_2+1}}{(u-1)^{\ell+1}} + \frac{c_2 + 8 c_1(u-2)}{4} \frac{u^{\Delta_3}}{(u-1)^{\ell+1}} \ , \\
f_4 &=& c_4 \frac{u^{\Delta_2}}{(u-1)^{\ell+1}} \ , \\
f_5 &=& 2c'_1 (u-2) \frac{u^{\Delta_2}}{(u-1)^{\ell+1}} + \frac{c_2 - 8 c_1 u}{4} \frac{u^{\Delta_3}}{(u-1)^{\ell+1}} \ , \\
f_6 &=& -2 c_2 \frac{u^{\Delta_3}}{(u-1)^{\ell+1}} + 2 c_4 \frac{u^{\Delta_2}}{(u-1)^{\ell+1}} \ .
\ee
\end{subequations}

From the defect OPE point of view, there are two defect three point functions which will contribute to 
$\langle F_{\mu\nu} \hat {\mathcal W}_2 \hat{\mathcal W}_3 \rangle$:
\be
\left \langle \hat \psi_-^{(-s)}({\bf x}_1) \hat {\mathcal W}_2({\bf x}_2) \hat {\mathcal W}_3({\bf x}_3)
\right\rangle &=& \frac{c_{\psi 23}}{w_{12}^s w_{23}^{\Delta_3 - \ell} \bar w_{12}^{s+1} \bar w_{13} \bar w_{23}^{\Delta_3 + \ell-1} } \ ,
\\
\left \langle \hat \phi_-^{(-s)}({\bf x}_1) \hat {\mathcal W}_2({\bf x}_2) \hat {\mathcal W}_3({\bf x}_3)
\right\rangle &=&  \frac{c_{\phi 23}}{w_{12}^{s+1} w_{13} w_{23}^{\Delta_3 - \ell-1} \bar w_{12}^{s}  \bar w_{23}^{\Delta_3 + \ell}} \ . 
\ee
To generate the bulk-defect-defect function, we then act on the first one with ${\mathcal B}^{(s)}_{\mu\nu}$ and the second with
${\mathcal D}^{(s)}_{\mu\nu}$.  Let us examine the action of these differential operators in the limits $|w_2|$ large and
$|w_3|$ large, in turn.   
In the limit $|w_3|$ large, the result should be $O(w_3^{- \Delta_3 +\ell} \bar w_3^{-\Delta_3 -\ell})$.
In contrast, because of the vanishing factors of $w_{13}$ and $\bar w_{13}$ in the denominators, in the limit
$|w_2|$ large, the result instead will be order $O(w_2^{-\Delta_3 + \ell-s-1} \bar w_2^{-\Delta_3 - \ell-s})$ from the
$\hat \psi_-^{(-s)}$ term and $O(w_2^{-\Delta_3 + \ell-s} \bar w_2^{-\Delta_3 - \ell-s-1})$ from the $\hat \phi_-^{(-s)}$ term.  

While the free field constraints give the same behavior in the large $|w_3|$ limit, they are instead less suppressed, 
$O(w_2^{-\Delta_3 + \ell-s} w_2^{-\Delta_3-\ell-s})$, in the large $|w_2|$ limit.  The only way to accommodate this difference is to 
set $c_4 = 0$.  A more detailed comparison reveals furthermore that $c_1' = 0$, $c_1 \sim c_{\phi 23}$ and 
$c_2 \sim 8 ( c_{\psi 23} +  c_{\phi 23}) $.

Finally, we invoke finiteness at $u=1$, which requires that $c_2 - 8 c_1 - 8 c'_1$ and $c_4 - 8 c_1 - 8 c'_1$ both vanish.
Compatibility with the defect OPE then sets $c_{\phi 23}$ and $c_{\psi 23}$ and the entire three-point function to zero.
One could imagine obtaining a consistent large $|w_2|$ scaling if the missing factors of $w_{13}$ and $\bar w_{13}$ were replaced with logarithms in the defect three point function.  Such a replacement, however, would be incompatible with conformal symmetry.  

The analysis for $s<0$ follows analogously.

\subsection{Case: $s=0$,  $\Delta_2 = \Delta_3$}

\begin{subequations}
\be
f_1 &=& \frac{1}{8 (\ell_2- \ell_3)} \biggl(
( \ell_3 - \ell_2 + 2 \Delta(u-1) - (2 + \ell_2 + \ell_3) u) f_2 - 2 u (u-1) f_2' \biggr) \ , \\
f_4 &=& - f_2 \ , \\
f_5 &=& \frac{1}{u (\ell_2 - \ell_3-2)}\biggl( (-4 \Delta(u-1) + (2 + \ell_2 +3 \ell_3) u) f_3 + 4(u-1)u f_3' \biggr) \ , \\
f_6 &=& -\frac{8}{u-1} (f_3 - f_5) \ .
\ee
\end{subequations}
The remaining second order ODEs for $f_2$ and $f_3$ can be solved to give
\be
f_2 &=& c_2 \frac{u^\Delta}{(u-1)^{1 + \frac{\ell_2 + \ell_3}{2}}} {}_2 F_1 \left( \frac{\ell_2 - \ell_3}{2}, \frac{\ell_3 - \ell_2}{2}, 1, \frac{1}{1-u} \right) \ , \\
f_3 &=& c_3 \frac{u^\Delta}{(u-1)^{ 1 + \frac{\ell_2 + \ell_3}{2}} } {}_2 F_1 \left( \frac{\ell_2 - \ell_3}{2}, \frac{\ell_3 - \ell_2 +2}{2}, 2, \frac{1}{1-u} \right) \ .
\ee

Matching to the defect OPE, we find that $c_2 \sim c_{\pi 23}$ while $c_3 \sim (2 - \ell_2 + \ell_3) c_{\tau 23}$.
Near $u= 1$, we find
\be
 \lefteqn{ \frac{1}{c_{\pi 23}} \langle F_{w \bar w}(x_1) \hat{\mathcal W}_2({\bf x}_2) \hat{\mathcal W}_3({\bf x}_3) \rangle  \sim
\frac{1}{c_{\tau 23} } \langle F_{z \bar z}(x_1) \hat{\mathcal W}_2({\bf x}_2) \hat{\mathcal W}_3({\bf x}_3) \rangle }  \\
&\sim&
(u-1)^{\frac{\ell_3 - \ell_2-1}{2}}
  \frac{ \Gamma(1+\ell_2-  \ell_3) }{\Gamma \left(1+ \frac{\ell_2 - \ell_3}{2} \right)^2  }
\pm (u-1)^{\frac{\ell_2-\ell_3-1}{2}} 
   \frac{ \Gamma(1+\ell_3 - \ell_2)}{\Gamma \left(1+ \frac{\ell_3 - \ell_2}{2} \right)^2  } \ ,  \nonumber
\ee
When $\Delta_2 = \Delta_3$, we have $(\ell_3 - \ell_2)/2 = h'_3 - h'_2 = h_2 - h_3$, giving us again the same double twist condition.

\subsection{Case: $s=0$, $h_2 = h_3$}

\begin{subequations}
\be
f_1 &=& - \frac{c_2}{8} (u-1)^{-1-\ell_3} u^{\Delta_3} \ , \\
f_2 &=& c_2 (u-1)^{-1 - \ell_3} u^{\Delta_3} \ , \\ 
f_3 &=& c_3 (u-1)^{-1-\ell_3} u^{\Delta_3} \ , \\ 
f_4 &=& c_4 (u-1)^{-1-\ell_2} u^{\Delta_2} \ , \\
f_5 &=& c_3 (u-1)^{-1-\ell_3} u^{\Delta_3} \ , \\
f_6 &=& c_6 (u-1)^{-1-\ell_2} u^{\Delta_2}- 8 c_3 (u-1)^{-1-\ell_3} u^{\Delta_3} \ .
\ee
\end{subequations}
In the case $\ell_2 \neq \ell_3$, we should then set either ($c_4$ and $c_6$) or ($c_2$ and $c_3$) to zero depending on whether
$\ell_2 - \ell_3$ is respectively positive or negative. 

However, such a choice is not consistent with the defect OPE.  Let us examine the defect OPE:
\be
\lefteqn{ \left \langle F_{z \bar z}(x_1) \hat {\mathcal W}^{(s)}_{(h_2, h'_2)}(x_2) \hat {\mathcal W}^{(-s)}_{(h_3, h_3')} (x_3) \right \rangle 
=} \\
&=&  -\frac{i}{2} {\mathcal D}  \left \langle \perpscalar ({\bf x}_1) \hat {\mathcal W}^{(s)}_{(h_2, h'_2)}(x_2) \hat {\mathcal W}^{(-s)}_{(h_3, h_3')} (x_3)  \right \rangle 
\nonumber \\
&=& -\frac{i}{2} \sum_{n=0}^\infty \frac{(-|z|^2)^n}{n!^2} \partial_{z_1}^n  \partial_{\bar z_1}^n
 \frac{c_{\perpscalarnohat 23} }{w_{12}^{h_1} w_{23}^{2h_2-h_1} w_{31}^{h_1} 
 \bar w_{12}^{h'_1+h'_2-h'_3} \bar w_{23}^{h'_2+h'_3-h'_1} \bar w_{31}^{h'_3 + h'_1 - h'_{2}} }\ .  \nonumber
\ee
We consider two separate limits, $w_2 \to \infty$ and $w_3 \to \infty$.  In the first case, we can write the result as
\be
\left \langle F_{z \bar z}(x_1) \hat {\mathcal W}^{(s)}_{(h_2, h'_2)}(x_2) \hat {\mathcal W}^{(-s)}_{(h_3, h_3')} (x_3) \right \rangle  
\to -\frac{i}{2} \frac{c_{\perpscalarnohat 23}}{w_2^{2 h_2} \bar w_2 ^{2  h'_2} w_{13} \bar w_{13}^{1+h'_3 - h'_2}} \left( 1 + \left| \frac{z}{w_{13}} \right|^2 \right)^{-1 + h_2'-h_3'} \ .
\ee
In the second, we get instead the same result with $2 \leftrightarrow 3$.  
This is to be compared with the constraint from the free field equations of motion, where in the first case, we find the same result but proportional to $c_6$ and in the second the same result but proportional to $c_3$.  
The results are mutually consistent only if $c_6 = 8 c_3 \sim c_{\perpscalarnohat 23}$.  

One can repeat the same exercise for the $w \bar w$ component in which case we find from the defect OPE a result proportional to the three point function coefficient $c_{\parscalarnohat 23}$ while from the free field constraints either $c_4$ or $c_2$. 
The results are consistent only if $c_4 = -c_2 \sim c_{\parscalarnohat 23}$.
This correlation between $c_6$ and $c_3$ in the first case and $c_4$ and $c_2$ in the second makes it impossible to render the solution finite at $u=1$ in general.  Thus we can rule out the situation where $h_2 = h_3$ (and $h'_2 \neq h'_3$).

\subsection{Case: $s=0$, $h'_2 = h'_3$}

\begin{subequations}
\be
f_1 &=& \frac{c_4}{8} (u-1)^{-1-\ell_3} u^{1+\Delta_2} + \frac{c_2}{8} (u-1)^{-\ell_2} u^{\Delta_3} \ , \\
f_2 &=& c_2 (u-1)^{-1-\ell_2} u^{\Delta_3} \ , \\
f_3 &=& c_3' (u-1)^{-1-\ell_3} u^{1 + \Delta_2} + c_3 (u-1)^{-\ell_2} u^{\Delta_3} \ , \\
f_4 &=& c_4 (u-1)^{-1-\ell_3} u^{\Delta_2} \ , \\
f_5 &=& -c_3' (u-2) (u-1)^{-1-\ell_3} u^{\Delta_2} - c_3 (u-1)^{-1-\ell_2} u^{\Delta_3} \ , \\
f_6 &=& -8 c_3' (u-1)^{-1-\ell_3} u^{\Delta_2} -8 c_3 (u-1)^{-1-\ell_2} u^{\Delta_3} \ .
\ee
\end{subequations}
In the case $\ell_2 \neq \ell_3$, we should then set either ($c_2$ and $c_3$) or ($c_3'$ and $c_4$) to zero depending on whether
$\ell_2 - \ell_3$ is respectively positive or negative. 

Similar to what happens in the $h_2 = h_3$ situation, this case is also incompatible with the defect OPE.
To be compatible, $c_2 = - c_4 \sim c_{\parscalarnohat 23}$ and $c_3 = -c'_3 \sim c_{\perpscalarnohat 23}$, which yields a solution which
diverges at $u=1$.  Thus we can rule out cases where $h'_2 = h'_3$ (and $h_2 \neq h_3$).

\subsection{Case: $s=0$, $\ell_2 = \ell_3$, $\Delta_2 = \Delta_3$}

The most general solution involves six integration constants:
\begin{subequations}
\be
f_1 &=& \frac{u^\Delta}{(u-1)^{\ell+1}} \left(- \frac{c_2}{8} + u c_1 \right) \ , \\
f_2 &=& \frac{c_2 u^\Delta}{(u-1)^{\ell+1}} \ , \\
f_3 &=& \frac{u^\Delta}{(u-1)^{\ell+1}} \left( c_3 + u c_3' \right) \ , \\
f_4 &=& \frac{c_4 u^\Delta}{(u-1)^{\ell+1}} \ , \\
f_5 &=& \frac{u^\Delta}{(u-1)^{\ell+1}} \left( c_3 - (u-2)  c_3' \right) \ , \\
f_6 &=& \frac{c_6 u^\Delta}{(u-1)^{\ell+1}} \ .
\ee
\end{subequations}

Eliminating the singularity at $u=1$ requires 
setting $c_1 = c_2/8$, $c_4 = c_2$, $c'_3 = -c_3$, and $c_6 = 0$.
The solution reduces to
\be
f_2 = f_4 = \frac{8 f_1}{u-1} = \frac{c_2 u^{\Delta}}{(u-1)^{\ell+1}} \ , \; \; \; f_5 = -f_3  = \frac{c_3 u^\Delta}{4(u-1)^\ell} \ , \; \; \; f_6 = 0 \ .
\ee

However this result is inconsistent with the defect OPE.  The defect OPE solution is $c_1 = c'_3 = c_6 = 0$ and
$c_2 = - c_4 \sim c_{\parscalarnohat 23}$ and $c_3 \sim c_{\perpscalarnohat 23}$, each branch of which gives a result which diverges at $u=1$, 
\be
 \left \langle F_{z \bar z}(x_1) \hat {\mathcal W}^{(s)}_{(h, h')}(x_2) \hat {\mathcal W}^{(-s)}_{(h, h')} (x_3) \right \rangle 
 =
c_{\perpscalarnohat 23}  \frac{ |w_{12}w_{13}|^2 - |z|^2 }{|w_{23}|^{2 \Delta+2} |z|^2}  \frac{1}{u(u-1)} \ ,
\ee
and a similar result for $F_{w \bar w}$ proportional to $c_{\parscalarnohat 23}$, 
suggesting that both
$c_{\perpscalarnohat 23} = c_{\parscalarnohat 23} = 0$ and ruling out the last special case $h_2 = h_3$ and $h'_2 = h'_3$.

\bibliographystyle{jhep}
\bibliography{Maxwellmonodromy.bib}

\providecommand{\href}[2]{#2}\begingroup\raggedright\begin{thebibliography}{10}

\bibitem{Herzog:2017xha}
C.~P. Herzog and K.-W. Huang, \emph{{Boundary Conformal Field Theory and a
  Boundary Central Charge}},
  \href{http://dx.doi.org/10.1007/JHEP10(2017)189}{\emph{JHEP} {\bf 10} (2017)
  189}, [\href{https://arxiv.org/abs/1707.06224}{{\tt 1707.06224}}].

\bibitem{Teber:2012de}
S.~Teber, \emph{{Electromagnetic current correlations in reduced quantum
  electrodynamics}},
  \href{http://dx.doi.org/10.1103/PhysRevD.86.025005}{\emph{Phys. Rev. D} {\bf
  86} (2012) 025005}, [\href{https://arxiv.org/abs/1204.5664}{{\tt
  1204.5664}}].

\bibitem{Kotikov:2013eha}
A.~V. Kotikov and S.~Teber, \emph{{Two-loop fermion self-energy in reduced
  quantum electrodynamics and application to the ultrarelativistic limit of
  graphene}}, \href{http://dx.doi.org/10.1103/PhysRevD.89.065038}{\emph{Phys.
  Rev. D} {\bf 89} (2014) 065038}, [\href{https://arxiv.org/abs/1312.2430}{{\tt
  1312.2430}}].

\bibitem{Dudal:2018pta}
D.~Dudal, A.~J. Mizher and P.~Pais, \emph{{Exact quantum scale invariance of
  three-dimensional reduced QED theories}},
  \href{http://dx.doi.org/10.1103/PhysRevD.99.045017}{\emph{Phys. Rev. D} {\bf
  99} (2019) 045017}, [\href{https://arxiv.org/abs/1808.04709}{{\tt
  1808.04709}}].

\bibitem{Gorbar:2001qt}
E.~V. Gorbar, V.~P. Gusynin and V.~A. Miransky, \emph{{Dynamical chiral
  symmetry breaking on a brane in reduced QED}},
  \href{http://dx.doi.org/10.1103/PhysRevD.64.105028}{\emph{Phys. Rev. D} {\bf
  64} (2001) 105028}, [\href{https://arxiv.org/abs/hep-ph/0105059}{{\tt
  hep-ph/0105059}}].

\bibitem{Kotikov:2016yrn}
A.~V. Kotikov and S.~Teber, \emph{{Critical behaviour of reduced QED$_{4,3}$
  and dynamical fermion gap generation in graphene}},
  \href{http://dx.doi.org/10.1103/PhysRevD.94.114010}{\emph{Phys. Rev. D} {\bf
  94} (2016) 114010}, [\href{https://arxiv.org/abs/1610.00934}{{\tt
  1610.00934}}].

\bibitem{Olivares:2021svj}
J.~A.~C. Olivares, A.~J. Mizher and A.~Raya, \emph{{Non-perturbative field
  theoretical aspects of graphene and related systems}},
  \href{https://arxiv.org/abs/2109.10420}{{\tt 2109.10420}}.

\bibitem{Hsiao:2017lch}
W.-H. Hsiao and D.~T. Son, \emph{{Duality and universal transport in
  mixed-dimension electrodynamics}},
  \href{http://dx.doi.org/10.1103/PhysRevB.96.075127}{\emph{Phys. Rev. B} {\bf
  96} (2017) 075127}, [\href{https://arxiv.org/abs/1705.01102}{{\tt
  1705.01102}}].

\bibitem{Hsiao:2018fsc}
W.-H. Hsiao and D.~T. Son, \emph{{Self-dual $\nu$ = 1 bosonic quantum Hall
  state in mixed-dimensional QED}},
  \href{http://dx.doi.org/10.1103/PhysRevB.100.235150}{\emph{Phys. Rev. B} {\bf
  100} (2019) 235150}, [\href{https://arxiv.org/abs/1809.06886}{{\tt
  1809.06886}}].

\bibitem{DiPietro:2019hqe}
L.~Di~Pietro, D.~Gaiotto, E.~Lauria and J.~Wu, \emph{{3d Abelian Gauge Theories
  at the Boundary}},
  \href{http://dx.doi.org/10.1007/JHEP05(2019)091}{\emph{JHEP} {\bf 05} (2019)
  091}, [\href{https://arxiv.org/abs/1902.09567}{{\tt 1902.09567}}].

\bibitem{Herzog:2018lqz}
C.~P. Herzog, K.-W. Huang, I.~Shamir and J.~Virrueta, \emph{{Superconformal
  Models for Graphene and Boundary Central Charges}},
  \href{http://dx.doi.org/10.1007/JHEP09(2018)161}{\emph{JHEP} {\bf 09} (2018)
  161}, [\href{https://arxiv.org/abs/1807.01700}{{\tt 1807.01700}}].

\bibitem{James:2021ggq}
A.~James, S.~Metayer and S.~Teber, \emph{{$\mathcal{N}=1$ supersymmetric
  three-dimensional QED in the large-$N_f$ limit and applications to
  super-graphene}},  \href{https://arxiv.org/abs/2102.02722}{{\tt 2102.02722}}.

\bibitem{KumarGupta:2019nay}
R.~Kumar~Gupta, C.~P. Herzog and I.~Jeon, \emph{{Duality and Transport for
  Supersymmetric Graphene from the Hemisphere Partition Function}},
  \href{http://dx.doi.org/10.1007/JHEP05(2020)023}{\emph{JHEP} {\bf 05} (2020)
  023}, [\href{https://arxiv.org/abs/1912.09225}{{\tt 1912.09225}}].

\bibitem{Gupta:2020eev}
R.~K. Gupta, A.~Ray and K.~Sil, \emph{{Supersymmetric graphene on squashed
  hemisphere}}, \href{http://dx.doi.org/10.1007/JHEP07(2021)074}{\emph{JHEP}
  {\bf 07} (2021) 074}, [\href{https://arxiv.org/abs/2012.01990}{{\tt
  2012.01990}}].

\bibitem{DiehlEisenrieglerArticle}
H.~W. Diehl and E.~Eisenriegler, \emph{Surface critical behavior of tricritical
  systems}, \href{http://dx.doi.org/10.1103/PhysRevB.37.5257}{\emph{Phys. Rev.
  B} {\bf 37} (1987) 5257}.

\bibitem{Prochazka:2020vog}
V.~Proch\'azka and A.~S\"oderberg, \emph{{Spontaneous symmetry breaking in free
  theories with boundary potentials}},
  \href{https://arxiv.org/abs/2012.00701}{{\tt 2012.00701}}.

\bibitem{Giombi:2019enr}
S.~Giombi and H.~Khanchandani, \emph{{$O(N)$ models with boundary interactions
  and their long range generalizations}},
  \href{http://dx.doi.org/10.1007/JHEP08(2020)010}{\emph{JHEP} {\bf 08} (2020)
  010}, [\href{https://arxiv.org/abs/1912.08169}{{\tt 1912.08169}}].

\bibitem{Giombi:2020rmc}
S.~Giombi and H.~Khanchandani, \emph{{CFT in AdS and boundary RG flows}},
  \href{http://dx.doi.org/10.1007/JHEP11(2020)118}{\emph{JHEP} {\bf 11} (2020)
  118}, [\href{https://arxiv.org/abs/2007.04955}{{\tt 2007.04955}}].

\bibitem{DiPietro:2020fya}
L.~Di~Pietro, E.~Lauria and P.~Niro, \emph{{3d large $N$ vector models at the
  boundary}},
  \href{http://dx.doi.org/10.21468/SciPostPhys.11.3.050}{\emph{SciPost Phys.}
  {\bf 11} (2021) 050}, [\href{https://arxiv.org/abs/2012.07733}{{\tt
  2012.07733}}].

\bibitem{Giombi:2021cnr}
S.~Giombi, E.~Helfenberger and H.~Khanchandani, \emph{{Fermions in AdS and
  Gross-Neveu BCFT}},  \href{https://arxiv.org/abs/2110.04268}{{\tt
  2110.04268}}.

\bibitem{Herzog:2020lel}
C.~P. Herzog and N.~Kobayashi, \emph{{The $O(N)$ model with $\phi^6$ potential
  in ${\mathbb R}^2 \times {\mathbb R}^+$}},
  \href{http://dx.doi.org/10.1007/JHEP09(2020)126}{\emph{JHEP} {\bf 09} (2020)
  126}, [\href{https://arxiv.org/abs/2005.07863}{{\tt 2005.07863}}].

\bibitem{Lauria:2020emq}
E.~Lauria, P.~Liendo, B.~C. Van~Rees and X.~Zhao, \emph{{Line and surface
  defects for the free scalar field}},
  \href{http://dx.doi.org/10.1007/JHEP01(2021)060}{\emph{JHEP} {\bf 01} (2021)
  060}, [\href{https://arxiv.org/abs/2005.02413}{{\tt 2005.02413}}].

\bibitem{Heydeman:2020ijz}
M.~Heydeman, C.~B. Jepsen, Z.~Ji and A.~Yarom, \emph{{Renormalization and
  conformal invariance of non-local quantum electrodynamics}},
  \href{http://dx.doi.org/10.1007/JHEP08(2020)007}{\emph{JHEP} {\bf 08} (2020)
  007}, [\href{https://arxiv.org/abs/2003.07895}{{\tt 2003.07895}}].

\bibitem{Herzog:2020bqw}
C.~P. Herzog and A.~Shrestha, \emph{{Two point functions in defect CFTs}},
  \href{http://dx.doi.org/10.1007/JHEP04(2021)226}{\emph{JHEP} {\bf 04} (2021)
  226}, [\href{https://arxiv.org/abs/2010.04995}{{\tt 2010.04995}}].

\bibitem{McAvity:1993ue}
D.~M. McAvity and H.~Osborn, \emph{{Energy momentum tensor in conformal field
  theories near a boundary}},
  \href{http://dx.doi.org/10.1016/0550-3213(93)90005-A}{\emph{Nucl. Phys. B}
  {\bf 406} (1993) 655--680}, [\href{https://arxiv.org/abs/hep-th/9302068}{{\tt
  hep-th/9302068}}].

\bibitem{McAvity:1995zd}
D.~M. McAvity and H.~Osborn, \emph{{Conformal field theories near a boundary in
  general dimensions}},
  \href{http://dx.doi.org/10.1016/0550-3213(95)00476-9}{\emph{Nucl. Phys. B}
  {\bf 455} (1995) 522--576},
  [\href{https://arxiv.org/abs/cond-mat/9505127}{{\tt cond-mat/9505127}}].

\bibitem{Billo:2016cpy}
M.~Bill\`o, V.~Gon\c{c}alves, E.~Lauria and M.~Meineri, \emph{{Defects in
  conformal field theory}},
  \href{http://dx.doi.org/10.1007/JHEP04(2016)091}{\emph{JHEP} {\bf 04} (2016)
  091}, [\href{https://arxiv.org/abs/1601.02883}{{\tt 1601.02883}}].

\bibitem{Lauria:2017wav}
E.~Lauria, M.~Meineri and E.~Trevisani, \emph{{Radial coordinates for defect
  CFTs}}, \href{http://dx.doi.org/10.1007/JHEP11(2018)148}{\emph{JHEP} {\bf 11}
  (2018) 148}, [\href{https://arxiv.org/abs/1712.07668}{{\tt 1712.07668}}].

\bibitem{Lauria:2018klo}
E.~Lauria, M.~Meineri and E.~Trevisani, \emph{{Spinning operators and defects
  in conformal field theory}},
  \href{http://dx.doi.org/10.1007/JHEP08(2019)066}{\emph{JHEP} {\bf 08} (2019)
  066}, [\href{https://arxiv.org/abs/1807.02522}{{\tt 1807.02522}}].

\bibitem{Guha:2018snh}
S.~Guha and B.~Nagaraj, \emph{{Correlators of Mixed Symmetry Operators in
  Defect CFTs}}, \href{http://dx.doi.org/10.1007/JHEP10(2018)198}{\emph{JHEP}
  {\bf 10} (2018) 198}, [\href{https://arxiv.org/abs/1805.12341}{{\tt
  1805.12341}}].

\bibitem{Kobayashi:2018okw}
N.~Kobayashi and T.~Nishioka, \emph{{Spinning conformal defects}},
  \href{http://dx.doi.org/10.1007/JHEP09(2018)134}{\emph{JHEP} {\bf 09} (2018)
  134}, [\href{https://arxiv.org/abs/1805.05967}{{\tt 1805.05967}}].

\bibitem{Bianchi:2021snj}
L.~Bianchi, A.~Chalabi, V.~Proch\'azka, B.~Robinson and J.~Sisti,
  \emph{{Monodromy defects in free field theories}},
  \href{http://dx.doi.org/10.1007/JHEP08(2021)013}{\emph{JHEP} {\bf 08} (2021)
  013}, [\href{https://arxiv.org/abs/2104.01220}{{\tt 2104.01220}}].

\bibitem{Dolan:2000ut}
F.~A. Dolan and H.~Osborn, \emph{{Conformal four point functions and the
  operator product expansion}},
  \href{http://dx.doi.org/10.1016/S0550-3213(01)00013-X}{\emph{Nucl. Phys. B}
  {\bf 599} (2001) 459--496}, [\href{https://arxiv.org/abs/hep-th/0011040}{{\tt
  hep-th/0011040}}].

\end{thebibliography}\endgroup

\end{document}